\documentstyle[12pt,epsf]{article}
\input{psfig}

\def\ap#1,#2,#3#4{           {\it Ann. Phys. (NY)\/ }{\bf #1} #2 (19#3#4)}
\def\apj#1,#2,#3#4{          {\it Astrophys. J.\/ }{\bf #1} #2 (19#3#4)}
\def\apjl#1,#2,#3#4{         {\it Astrophys. J. Lett.\/ }{\bf #1} #2 (19#3#4)}
\def\app#1,#2,#3#4{          {\it Acta Phys. Polon.\/ }{\bf #1} #2 (19#3#4)}
\def\com#1,#2,#3#4{          {\it Comm. Math. Phys.\/ }{\bf #1} #2 (19#3#4)}
\def\ib#1,#2,#3#4{           {\it ibid.\/ }{\bf #1} #2 (19#3#4)}
\def\nat#1,#2,#3#4{          {\it Nature (London)\/ }{\bf #1} #2 (19#3#4)}
\def\np#1,#2,#3#4{           {\it Nucl. Phys.\/ }{\bf B#1} #2 (19#3#4)}
\def\npps#1,#2,#3#4{         {\it Nucl. Phys. B (Proc. Suppl.)\/ }{\bf B#1}
                             #2 (19#3#4)}
\def\plb#1,#2,#3#4{          {\it Phys. Lett.\/ }{\bf B#1} #2 (19#3#4)}
\def\pla#1,#2,#3#4{          {\it Phys. Lett.\/ }{\bf A#1} #2 (19#3#4)}
\def\prd#1,#2,#3#4{          {\it Phys. Rev.\/ }{\bf D#1} #2 (19#3#4)}
\def\prep#1,#2,#3#4{         {\it Phys. Rep.\/ }{\bf #1} #2 (19#3#4)}
\def\prl#1,#2,#3#4{          {\it Phys. Rev. Lett.\/ }{\bf #1} #2 (19#3#4)}
\def\pro#1,#2,#3#4{          {\it Prog. Theor. Phys.\/ }{\bf #1} #2 (19#3#4)}
\def\rmp#1,#2,#3#4{          {\it Rev. Mod. Phys.\/ }{\bf #1} #2 (19#3#4)}
\def\sp#1,#2,#3#4{           {\it Sov. Phys.-Usp.\/ }{\bf #1} #2 (19#3#4)}

\def\zp#1,#2,#3#4{           {\it Zeit. fur Physik\/ }{\bf C#1} #2 (19#3#4)}
\def\md{\mbox{$M_{D}$}}
\def\ra{\rightarrow}

\def\herw{{\sc  herwig}}

\def\pyt{{\sc  pythia}}

\def\ttb{\mbox{$t\bar{t}$}}

\def\vereq#1#2{\lower3pt\vbox{\baselineskip1.5pt \lineskip1.5pt}}

\def\beq{\begin{equation}}
\newcommand{\Ht}{\mbox{$H_{T}$}}
\newcommand{\et}{\mbox{$E_{T}$}}
\newcommand{\pt}{\mbox{$P_{T}$}}

\newcommand{\Wen}{\mbox{$W \rightarrow\ e \nu$}}
\newcommand{\Wtn}{\mbox{$W \rightarrow\ \tau \nu$}}
\newcommand{\Wmn}{\mbox{$W \rightarrow\ \mu \nu$}}
\newcommand{\Zee}{\mbox{$Z \rightarrow\ e e$}}

\newcommand{\bspace}{\!\!\!\!}
\newcommand{\met}{\mbox{$E{\bspace}/_{T}$}}
\newcommand{\niso}{\mbox{$N_{trk}^{iso}$}}

\def\stilde{\widetilde}
\newcommand{\sq}{\stilde{q}}

\newcommand{\gls}{\stilde{g}}
\newcommand{\msq}{\mbox {$m_{\stilde{q}}$} }
\newcommand{\mgls}{\mbox {$m_{\stilde{g}}$} }

\def\eeq{\end{equation}}
\def\bea{\begin{eqnarray}}
\def\beaa{\begin{eqnarray*}}
\def\eea{\end{eqnarray}}
\def\eeaa{\end{eqnarray*}}
\def\bq{\begin{quote}}
\def\eq{\end{quote}}

\def\gappeq{\mathrel{\rlap {\raise.5ex\hbox{$>$}}
{\lower.5ex\hbox{$\sim$}}}}
\def\lappeq{\mathrel{\rlap{\raise.5ex\hbox{$<$}}
{\lower.5ex\hbox{$\sim$}}}}
\def\sm{Standard Model}

\def\PRL{{ Phys.Rev.Lett.} }

\def\PRL{Phys.~Rev.~Lett.}

\begin{document}
\cleardoublepage
\title{Collider Experiment:\\ Strings, Branes and Extra Dimensions}

\author{Maria Spiropulu\\
Enrico Fermi Institute\\
5640 S. Ellis Ave.\\
Chicago IL, USA}

\maketitle\abstract{Selected topics showcasing the exploration for new physics using colliders; presented at TASI 2001.}

\thispagestyle{empty}
\catcode`@=11\def\tableofcontents{\subsection*{\contentsname
\@mkboth{\uppercase{\contentsname}}{\uppercase{\contentsname}}}%
\@starttoc{toc}}
\catcode`@=12
\setcounter{tocdepth}{2}
\tableofcontents
\cleardoublepage
\setcounter{page}{1}
\catcode`@=11\def\section{
\@startsection {section}{1}{\z@}{19pt plus1ex minus
 .2ex}{12pt plus1ex minus.2ex}{\reset@font\large\bf}}
\section{Preface}

The subject of the Theoretical Advanced Study Institute summer 2001 school was 
``STRINGS, BRANES and EXTRA DIMENSIONS''.  
Although not too many years ago a course on collider physics  would had been an unorthodox 
inclusion to the curriculum of a such titled school, today there are attempts to connect 
strings, branes and extra dimensions with experiment. Indeed string 
phenomenology is not an oxymoron, to the delight of all. I was particularly 
content to see young theorists spending time calculating the beam power
in a future linear electron accelerator and the dollar sum necessary to pay the
electric bill in a year. I was also very impressed that there were students 
who already knew what a trigger is and why we use it. 

It took at least a generation of inspired physicists and also math experts 
to move from field theories to string theory, and a concurrent of technowizzes and
inquiring minds to
discover and measure with exquisite precision the theory
describing nature at the most fundamental length scales yet explored: 
the Standard Model. Although
it seemed briefly that theory and experiment are fast growing apart we witness today 
a remarkable exchange between the two, and a tendency to almost believe 
that by putting separate small bricks of knowledge together, one day preferably soon, 
the complete edifice of nature will be exactly blueprinted and raised. 


The lectures were organized in three sections. The first one is devoted to accelerators.
There is no doubt that these machines gave birth to the field we call High Energy Physics
and filled in turn the Particles and Fields data-books. And there is no substitute for these
machines. They are evolving, with  the purpose of exploring the physics at the highest energy reachable or equivalently 
the physics at the shortest length scale. The second is an overview of
the kind of experiments that lead us from atoms all the way down to quarks and 
a brief summary of the particle physics jargon used when we report results. 
And the third is analysis examples and interpretation of
results when looking for new physics. Since the school's interests is
on string theory I will give a supersymmetry search example and an
example of the search for extra dimensions with collider data.

\catcode`@=12
\cleardoublepage
\section{Accelerators}
\label{chap:accelerators}
	There are at least 10,000 accelerators in the world, most of
	them put to action in solving every day problems. A world-wide
	search by W.H. Scharf and O.A. Chomicki \cite{medacc} 
	reports 112 accelerators
	of more than 1 GeV. A third of these are used in high energy
	physics research. The rest are mostly synchrotron light
	sources. About 5000 accelerating machines of lower energy are
	used in medicine (radiotherapy, biomedical research and isotope
	production). About as many are used in industry, usually for
	ion implantation and surface physics. The latest sterilization
	via irradiation made the news in the post September 11 time 
	when all the letters to Washington were collected to be
	irradiated for fear of anthrax. A great review of 
	accelerators for medical applications has been written by Ugo Amaldi 
	\cite{amaldi}.

	We are familiar with the use of high energy lepton
	and hadron accelerators for particle physics. Present
	investigations are focused on the search for the Higgs boson,
	neutrino oscillations, heavy quark physics,
	the production of supersymmetric particles and
	even the geometry and geography of spacetime. There is little doubt
	that this type of research is connected intimately with cosmology in 
	re-creating particles and interactions from the first instant of the
	Big Bang through the era when nuclei were formed by the more 
	fundamental particles. The data
	accumulated from high energy particle collisions are essential
	in formulating and stimulating cosmological models and in
	helping understand the origin of the dark matter and perhaps
	even the dark energy in the universe.

	I would like at this point to remark that it seems to me that
	when the director of the Office of Science and Technology Policy, Dr.
	Marburger, wrote ~\cite{apsnewsdec02}  
	that at `` some point 
	we will have to stop building accelerators'' he meant it in 
	a similar way that NASA will stop building space shuttles. This is
	to say that new technologies will necessarily  be employed in 
	endeavors  of tremendous magnitude such as the exploration 
	of the universe in all scales; the exploration itself will not cease.

\begin{table}
\begin{tabular}{lc} \hline 
Accelerators& Number in use\\ \hline 
(1) High-energy more than 1GeV& $~$112 \\ 
(2) Radiotherapy&  $>$4000\\ 
(3) Research/Biomedical Research & $\sim$800 $\sim$5000\\ 
(4) Medical radioisotope production & $\sim$ 200 \\ 
(5) Industry		 	  &  $\sim$1500\\
(6) Ion implanters		  & $>$2000\\
(7) Surface modification centers and research & $\sim$1000\\
(8) Synchrotron radiation sources & 50\\
total (in 1995) $\sim$10000
\end{tabular}
\caption[Accelerators in action]{Statistics of operating accelerators (1995).}
\label{sec1:accel}
\end{table}

\subsection{DC Accelerators}

	In the beginning of the century there was chemistry, philosophy (cosmic ray 
	research was being published in journals of philosophy), and by 1932
	we had nuclear physics when the neutron was discovered by Chadwick.
	Three more particles were known by then; the electron
 	(J.J. Thompson in 1897 passed electrons
	through crossed E and B fields, measured the velocity and then e/m of
	the electron), the proton (Rutherford, 1913 - he called the proton 
	the hydrogen atom nucleus) and the photon (Max Planck, black body
	radiation in quanta of $h\nu$, Einstein photoelectric effect, Compton
	X-ray scattering from electrons as if particles. The photon was 
	actually named by a chemist.)
 
	During the first part of the century natural radioactivity and 
	cosmic rays were the source of energetic particles for atomic 
	physics research. In 1906 Rutherford bombarded a mica 
	sheet with alpha particles from a natural radioactive 
	source (Rutherford scattering).
	Natural sources are limited in energy and intensity. In 1928 Cockroft
	and Walton started thinking about building an accelerator to use at the
	Cavendish Laboratory. In 1932 the apparatus was finished and used to
	split lithium nuclei with 400 keV protons. The measurement of the binding
	energy  in this experiment 
	provided the first experimental verification of Einstein's 
	mass-energy relationship (known since 1905). This was the start of
	particle accelerators for nuclear research. The very first ones 
	were direct voltage accelerators like the Cockroft and Walton
	(rectifier generator up to 1 MV), the Van de Graff generator (up to 
	10 MV in high-pressure tank containing dry nitrogen or freon
	to avoid sparking) and the tandem electrostatic accelerator (the
	accelerator is known as tandem because the ions, at the beginning negative, 
	undergo a double acceleration: they are attracted by the positive 
	central electrode, pass in a cleaner who makes them 
	positive, and they are then pushed back by the electrode; 
	up to maybe 35 MV, Vivitron Strasbourg, in operation now at 20 MV).

\subsection{AC Accelerators}
	Ising in 1924 proposed the first particle accelerator that 
	would give the particles more energy than the 
	maximum voltage in the system. He proposed an electron linear
	accelerator with drift tubes (but did not build it). In 1928 Wideroe 
	used an alternating 25 kV voltage with 1 MHz frequency applied over
	two gaps and produced 50 keV potassium ions. The Wideroe type
	linac comprises a series of conducting drift tubes. Alternate drift
	tubes are connected to the same terminal of the RF generator. 
	The frequency is such that when a particle goes through the gap 
	it sees the accelerating field and when the field becomes
	decelerating the particle is shielded inside the drift tube.
	As the particle gains in energy and velocity the structure periods 
	must be longer in order to be in sync. At very high frequencies 
	(so that the structure does not become inconveniently long) 
	the open drift-tube scheme needs to be enclosed to form a cavity or
	series of cavities.

	If one applies Ising's resonant principle in a homogeneous magnetic field 
	the particle would be bent back to the same RF gap twice for each 
	period. This is Lawrence's and Livingston's fixed-frequency {\it cyclotron}
	(the initial was less than one foot in diameter and could accelerate
	protons to 1.25 MeV). The resonance condition in the cyclotron is
	obtained by choosing the RF period equal to the cyclotron period,
	 which is independent of the particle velocity and the orbit radius ;
	it depends on the $q/m$ ratio and the magnetic field. Particles that
	pass the gap near the peak of the RF voltage would continue to do so
	every half turn moving in ever increasing half-circles (spiraling) 
	until they reach  the edge of the magnetic field or until they become
	relativistic and slip back with respect to the gap voltage. 
	The intrinsic limit was confronted in the late thirties at about 25 MeV 
	for protons and  50 MeV for deuterons and alpha particles.
	The cyclotron consisted of two ``D'' shaped
	regions in vacuum, called {\it dees} with a gap separating
	them and a magnetic field applied perpendicular to the dees. 
	As the proton beam crosses the gap it experiences an 
	electric field which gives the proton a kick and increases its energy. It
	gets an energy increase every time it crosses the gap. 
	The frequency of the applied electric field  is constant, while the
	radius of the proton beam keeps increasing.

	{\it Synchro-cyclotron} -or the frequency-modulated cyclotron- was the remedy
	for the relativistic limit in which the revolution frequency decreases
	with increasing energy, and the frequency of the accelerating voltage
	must also be correspondingly decreased.  Although this is necessary, it
	in not sufficient to maintain sync, because the natural energy spread in
	a bunch of relativistic ions causes a spread in their cyclotron
	frequencies, and thus longitudinal focusing is required to maintain the 
	``bunch''. The problem was overcome by McMillan and Veksler who
	discovered the principle of phase stability in 1944. The effect of 
	phase stability is that a bunch of charged particles with an energy
	spread can be kept bunched throughout the acceleration cycle by
	injecting them at a suitable phase of the RF cycle. Synchro-cyclotrons
	can be used to accelerate protons up to 1 GeV. The higher energy is
	obtained at the expense of intensity (number of particles in the bunch)
 	since the  pulsed beam has less intensity compared to a continuous beam. 
	To achieve transverse stability of the beam the field should be decreasing 
	with radius according to an inverse  power law. At Berkeley they found
	(the hard way) that the magnetic field has to decrease
	slightly with increasing orbit radius to prevent the particles from getting lost.

	For electrons the cyclotron is useless as they are very quickly
	relativistic.
	The solution was the  {\it betatron},  a  device that has found applications in
	laboratories and hospitals.  It was conceived again by Wideroe; Kernst built
	the first betatron in 1941 and Kernst and Serber published a paper
	on ``betatron oscillations'' the same year. In 1950 Kernst built the
	world's largest betatron. In a betatron, the electromagnet is
	powered with an AC current at 50 to 200 Hz. The magnetic field guides
	the particles in a circular orbit, but because it is a changing
	magnetic field, it induces a  circumferential voltage which accelerates
	the particles. The guide field was carefully shaped and given a
	radial gradient in order to provide vertical and horizontal beam stability.
	If the electron path is to remain at constant radius the magnetic
	field must increase as the electron energy increases. The increasing
	field results in increasing magnetic flux through the orbits which
	then induces the force that increases the energy of the electrons. 
	The magnetic flux through the orbit must be twice the bending field 
	in order to keep the beam at the same radius. 

	The betatron was soon replaced by the {\it synchrotron} which is an
	accelerator that combines the properties of the cyclotron and those
	of the betatron. McMillan and Veksler already discusses the idea of
	synchronous acceleration in their cyclotron papers.  The first
	synchrotron (Cosmotron) was a 3 GeV proton accelerator built in 1952
	at  Brookhaven National Laboratory (BNL).
	The machine had straight sections and a guide field 
	similar to the betatron with a bending
	field to keep the particles on a circular orbit and appropriate radial
	gradient to achieve vertical and horizontal stability. Acceleration was
	achieved by RF voltage at the revolution frequency. As the particle
	energy increases the field is also increased at a rate that keeps the
	particles in approximately the same orbit at all energies. This means
	that the RF voltage frequency should be increasing (compare with
	cyclotron).  Many large synchrotron machines followed the Cosmotron.
	The Tevatron at Fermilab is one of them, accelerating protons and
	antiprotons to approximately 1000 GeV. 
%
	
        In all accelerators build before 1952 the 	
	transverse stability of the beam depended on what is now called
	{\it weak focusing} in the magnet system.  The guide field decreases
	slightly with increasing radius in the vicinity of the particle orbit 
	and this gradient is the same all around the circumference of the
	magnet. The tolerance on the gradient is severe and sets limits to
	such an accelerator (believed to be around 10 GeV in the early 
	fifties). The aperture needed to contain the beam becomes
	very large and the magnet becomes very bulky and costly. The invention
	of the {\it alternating-gradient} principle by Christofilos and
	independently by Courant, Livingston and Snyder in 1952 changed 
	this picture. As a matter of fact, machines in the range of 10-100 TeV
	seem already technically possible. The limitation is the cost.
	Alternating-gradient or ``strong''
        focusing is directly analogous to a result in 
	geometrical optics, that the combined focal length of two
	appropriately spaced lenses of equal strength will be overall focusing
	when one lens is focusing and the other defocusing. 
	Such a system remains focusing for quite some range of
	the focal length values of the two lenses. A quadrupole lense
	focuses on one plane and defocuses on the orthogonal plane.
	An appropriate arrangement of quadrupoles can be altogether focusing 
	in both planes.  Structures based on this principle are 
	called {\it alternating gradient (AG) structures}.{\footnote{
        The first such machine, the Brookhaven AGS, is still
        in operation today. The AGS was used to discover the 
	J/$\Psi$
	(Nobel 1976 S. Ting), 
	CP violation in the Kaon system (Nobel 1980 J. Cronin and V. Fitch)
	and the muon neutrino
        (Nobel 1988 J. Steinberger, L. Lederman, M. Schwartz).}}

	Although the circular machines took over for some time, the linear
	machines were revived after WW II with advances in ultra-high
	frequency technologies. At Berkeley a proton linac was built
	by Alvarez (1946) who employed the war-developed radar technology 
	and enclosed the entire drift tube structure in a resonant cavity
	to reduce losses.  Since then this type of accelerator has
	been widely used as an injector (with injection energies reaching 200 MeV) for
	large proton and heavy-ion synchrotrons (e.g.
        the Fermilab linac and the HERA
	linac at Deutches Electronen Synchrotron (DESY) in Hamburg,
	Germany). The largest proton linear accelerator today is the
	the 800 MeV proton linac at the Los Alamos Neutron Science Center
	(LANSE).
 	The largest electron linear accelerator in operation is 
	the 50 GeV linac at the Stanford Linear Accelerator Center (SLAC).
	Linear accelerators, like betatrons previously, have become
	very popular in fields outside particle physics (e.g materials science,
	biomedical research, and medicine).  
\begin{figure}
\centerline{
\psfig{figure=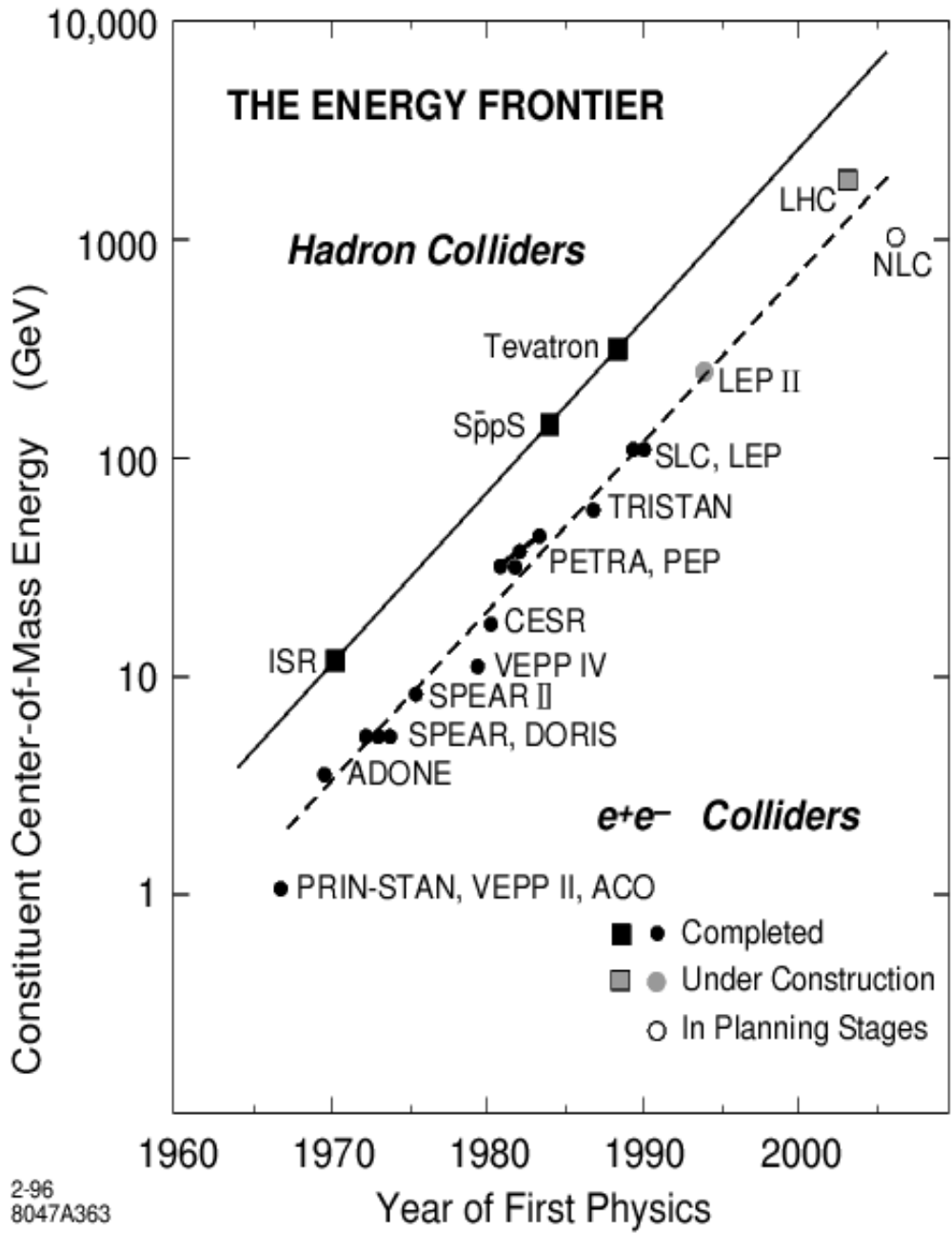}}
\caption[The energy frontier]
{The effective constituent energy of existing and planned colliders and the
year of first physics results from each (after \cite{nlc}).}	
\label{living}
\end{figure}

	An increase in beam energy of about 1.5 orders of magnitude per decade
	is illustrated by the Livingston chart shown in Figure ~\ref{living}.
	While the energy increases by 8 orders of magnitude the cost per GeV of a
	typical accelerator has been drastically reduced. 
	But what is most amazing is that while each type of acceleration is coming to
	saturation relatively fast, there is an advance in technology
	that allows a different idea on acceleration to kick in. 
	Examples: the invention of the alternating gradient focusing
	in the fifties, the {\it colliding beams} in the sixties, superconducting 
	magnet technology and stochastic cooling in the seventies and eighties. 


\subsection{Colliders/Storage Rings}
	In high energy  physics experiments, the type and number 
	of particles brought into collision and
	their center of mass energy characterize an interaction.	
The center of mass energy $E_{CM}$ made available when a synchrotron beam 
	of energy $E_b$ hits a fixed target is approximately
	\begin{equation}
	\sqrt{s}=E_{CM}=(2m_{target}E_{b})^{1/2}
	\end{equation}
	The center of mass energy when colliding head-on two beams of energy $E_{b}^{\prime}$ is
	\begin{equation}
	\sqrt{s}=E_{CM}=2E_{b}^{\prime}
	\end{equation}

	The purpose of storage rings is to make head-on collisions possible 
	with a useful interaction rate. 
	At the Tevatron the maximum proton energy is about  
 	1000 GeV.
        In the  $p\bar p$ collider  mode the center of mass energy is 
	$\sqrt{s}=E_{CM}=2000$ GeV.
	In the fixed target mode it is $\sqrt{s}=E_{CM}=41$ GeV.

	The fixed target collisions  
	produce a variety of secondary particles from the target that can  be collected 
	and focused into secondary beams. Also in fixed target collisions the
	achieved luminosities are extremely high.
	The reaction rate $R$ is $R=L~\times~ \sigma$, where $L$ is the luminosity and $\sigma$ is the cross section.	
	This implies that $L=(N_{beam}/\rm{sec})(N_{target}/\rm cm^{2})$. 
	Luminosity is measured 
	usually in cm$^{-2}$sec$^{-1}$. A fixed target experiment can have 
	luminosity as high as $10^{37}$ cm$^{-2}$sec$^{-1}$.  
	In a high luminosity collider experiment the  luminosity is of the order of 
	$10^{32}$  cm$^{-2}$sec$^{-1}$. 

	What is the limitation on the energy for an accelerator storage ring?
	For proton accelerators it is the maximum magnetic field to bend the particles
	($R_{ring}(m)=P(GeV)/0.3B(Tesla)$).
	For electron storage rings it is the energy lost to synchrotron radiation per revolution,
	$U\approx 0.0885E(GeV)^{4}R(m)^{-1}$. (For relativistic proton beams
	this is $\approx 7.79\time 10^{-15}E(\rm{GeV})^{4}R(\rm{m})^{-1}$).
	The largest electron-positron storage ring was the 27 km LEP 
	ring at CERN in Geneva. 
	It started running  at a beam energy of about 50 GeV:
	this meant that 200 MeV per turn should be supplied by the RF to make up
	 for the synchrotron radiation loss. 	When it ran at 100 GeV, a factor of two higher in energy,  the 
	energy loss per turn was 3.2 GeV, a factor of 16 increase in losses! 
	A linear electron-positron accelerator (such as the SLC, the Stanford Linear Collider) 	
	avoids the problem of synchrotron radiation. It is expected that all future
	electron-positron machines at higher energy than LEP will be linear.

	Note that a monoenergetic proton (antiproton) beam is equivalent
	to a wide-band parton beam (where parton$\equiv$quarks, antiquarks, gluons),
	described by momentum distribution $dn_{i}/dp$ 
	(structure function, $i$ specifies the parton type, i.e.
	$u,d,g,\bar{u},\bar{d}...$). Proton structure functions 
	are measured in {\it deep inelastic scattering} experiments. 

	Some of the advantages of hadron collisions include the
	simultaneous study of a wide energy interval, therefore there is 
	no requirement for precise tuning of the machine energy; the
	greater variety of initial state quantum numbers e.g. $u+\bar{d} \rightarrow
	W^{+},~ \bar{u}+d \rightarrow W^{-}$; the fact that the maximum energy
	is much higher than the maximum energy of $e^{+}e^{-}$ machines; 
	and finally that
hadron collisions are the only way to study parton-parton collisions 
	(including gluon-gluon). Some of the disadvantages are the huge cross sections
	for uninteresting events; the multiple parton collisions in the same hadron 
	collision that result in complicated final states; and that the center-of-mass frame of the
	colliding partons is not at rest at the lab frame.

	Table~\ref{accel} lists some present, historic and proposed colliders.

\begin{table}
\label{accel}
\begin{center}
\begin{tabular}{|l|l|l|l|l|}\hline\hline
\multicolumn{5}{c} {Proton Fixed Target Accelerators}  \\ \hline
Name & Lab & Location & GeV & status \\ \hline
PS & CERN & Geneva& 28& inactive\\ \hline
AGS & BNL & Long Island & 32& active\\ \hline
SPS & CERN & Geneva & 450 & active\\ \hline 
Tevatron & FNAL & Chicago & 1000 & active\\ \hline
\multicolumn{5}{c} {Electron Fixed Target Accelerators}  \\ \hline
LINAC & SLAC & Stanford & 50 & active\\ \hline
\multicolumn{5}{c} {Proton Colliding Beam Machines}  \\ \hline
ISR & CERN & Geneva & 31$p \times$ 31$p$& inactive \\ \hline
S$p\bar p$S & CERN & Geneva & 310$p \times$ 310$\bar p$& inactive \\ \hline
Tevatron & FNAL & Chicago & 1000$p \times$ 1000$\bar p$ & active\\ \hline
SSC & SSCL & Dallas & 20000$p \times$ 20000$\bar p$ & canceled \\ \hline
LHC  & CERN & Geneva & 7000$p \times$ 7000$p$& 2007 \\ \hline
\multicolumn{5}{c} {Electron Colliding Beam Machines}  \\ \hline
SPEAR & SLAC & Stanford & 4$e^{-}\times~ 4e^{+}$& inactive\\ \hline
DORIS & DESY & Hamburg & 5$e^{-}\times~ 5e^{+}$& inactive\\ \hline
BES & BES & Beijing  & 4$e^{-}\times~ 4e^{+}$& active\\ \hline
CESR & Cornell & Ithaca & 8$e^{-}\times~ 8e^{+}$& active\\ \hline
PEP & SLAC & Stanford  & 15$e^{-}\times~ 15e^{+}$& inactive\\ \hline
PEPII & SLAC & Stanford  & 9$e^{-}\times~ 3e^{+}$& active\\ \hline
PETRA & DESY & Hamburg &  23$e^{-}\times~ 23e^{+}$& inactive\\ \hline
TRISTAN & KEK & Tokyo  & 30$e^{-}\times~ 30e^{+}$& inactive\\ \hline
KEK B & KEKB & Tokyo  & 8$e^{-}\times~ 3.5 e^{+}$& active\\ \hline
SLC & SLAC & Stanford  & 50$e^{-}\times~ 50e^{+}$& active\\ \hline
LEPII & CERN & Geneva  & 200$e^{-}\times~ 200e^{+}$& inactive\\ \hline
\multicolumn{5}{c} {Electron-Proton  colliding Beam Machines}  \\ \hline
HERA & DESY & Hamburg & 30$e^{-}\times~ 820p$  & active\\ \hline
\multicolumn{5}{c} {Future Electron-Positron Collider Study Collaborations} \\ \hline
{JLC} & {-} & {KEKB} & (stage I) 125$e^{-}\times~125e^{+}$& R\&D \\ \hline
{NLC} & {-} & {SLAC} & 400$e^{-}\times~400e^{+}$ & R\&D \\ \hline
{TESLA} & {-} &{DESY} & 400$e^{-}\times~400e^{+}$ & R\&D \\ \hline 
{CLIC} & {-}  &{CERN} & 1500$e^{-}\times~1500e^{+}$ & R\&D  \\ \hline
\multicolumn{5}{c} {Future Hadron Collider Study Collaborations} \\ \hline
VLHC & FNAL & Chicago & 40000$p \times$ 40000$ p$ & R\&D \\ \hline 
\end{tabular}
\end{center}
\caption{Present, historic(including cancelled)  and proposed colliders}  
\end{table}

\subsection{Future Colliders}

We saw that the highest energy circular electron machine has been LEP.
In order to go any higher in energy in electron-positron collisions, 
 we need to build the collider on a straight line.
The luminosity in terms of the configuration of the beams
is  $L=\frac{f_{coll}N^{2}}{S}$ where $f_{coll}$ is the number of bunch
collisions per second, $N$ is the number of particles in a bunch and $S$ is the beam cross-sectional area at the collision point. In a linear collider clearly $f_{coll}$ is much smaller than 
in a circular collider. The number of particles $N$ in a bunch is similar or smaller.
Therefore to achieve worthwhile luminosities it is 
imperative that the beam emittance $S$ be much smaller.
A linear collider needs very high accelerating gradient 
to keep the site length reasonable and very high power efficiency 
to keep the cost under control. As we noted the beams need to be 
generated with extremely small emittance which has to be preserved during acceleration.
At the collision point the beams have to be tightly focused.
Small emittances can be achieved by 
preparing the beam through what is called a damping ring and making use of the 
synchrotron radiation. The acceleration is achieved by microwave cavities
which can be superconducting or normal conducting, and the 
frequency choice is based on the achievable accelerating 
gradient. There are
at least two different designs under development: The NLC/JLC 
{\footnote{Next Linear Collider, Japan Linear Collider}}
design that uses normal conducting copper cavities at low frequency
(11.424 GHz or X-band) which is also referred to as {\it warm} design, 
and the TESLA {\footnote{Tera Electronvolt Superconducting Linear Accelerator}}
design that uses supeconducting niobium cavities, a 
technology that is power consumption effective (there is extremely 
small power dissipation in the cavity walls) but thought to be 
very expensive until recently. This is referred to as {\it cold} design
because the cavities need to be kept at very low temperatures.
Linear collider accelerating structures can also be used to drive 
free electron lasers, with important applications
for chemistry, materials science, plasma research
and life sciences research.  

A post-LHC and probably post-LC collider that accelerates 
and collides electron and positron
beams with center of mass 3-5 TeV is envisioned in  the 
CLIC{\footnote{Compact Linear Collider}} design. It is based on the two-beam acceleration  method 
in which the RF power for sections of the main linac 
is extracted from a secondary, low-energy, high-intensity 
electron beam running parallel to the main linac.

The next hadron collider proposed is the VLHC {\footnote{Very Large Hadron Collider}}. 
A design study group
has developed the basic parameters, technology/construction 
challenges and cost  of a proton-proton 
collider with center of mass energy greater than 30 TeV, 
that would allow the eventual operation of a collider with
center of mass energy greater than 150 TeV in the same tunnel.
The machine would involve two phases, one with low-field
magnets and one with high-field magnets for the energy upgrade
in the same 233 km ring. 

It is not included in the table, but extensive study and
research is geared towards the 
feasibility and potential of high energy 
high luminosity muon colliders operating at a 
center-of-mass energy in the range 100 GeV - 4 TeV.
The high intensity muon source needed for muon 
colliders can also be used to feed a muon storage 
ring neutrino source (neutrino factory).
The enthusiasm in recent years for a muon collider is due to ionization cooling 
that allows for very bright muon 
beams. 

One lesson from the Livingston plot is that a new 
technology can push the collision energy and drive 
high energy research.  Indeed accelerator technology is also moving 
towards alternative directions.  High-gradient plasma wakefield 
acceleration is one. The plasma-wave acceleration process 
is dramatically strong and may lead
to very high energy beams with reasonable size accelerators.

\cleardoublepage
\section{From atoms to quarks, {\it Rutherford redux}}
	The kind of experiments we perform to probe 
	the structure of matter belong to three categories: scattering, 
	spectroscopy and break-up experiments. 
	The results of such experiments brought us
	from the atom to the nucleus, to hadrons, to quarks.

	Geiger and Marsden reported in 1906 the
	measurements of how  $\alpha$ particles 
	(the nuclei of He atoms) were deflected 
	by thin metal foils. They wrote ``it seems 
	surprising that some of the $\alpha$ particles,
	as the experiment shows, can be turned within 
	a layer of 6$\times 10^{-5}$ cm of gold through
	 an angle of 90$^{\circ}$, and even more''.
	Rutherford later said he was amazed as if he had seen a bullet
	bounce back from hitting a sheet of paper. 
	It took two years to find the explanation. 
	The atom was known to be neutral and to be containing negatively
	charged electrons of mass very much less than the mass of the atom. So 
	how was the positively charged mass distributed within the atom?
	Thompson had concluded that the scattering of the 
	$\alpha$ particles was a result 
	of ``a multitude of small scatterings
	by the atoms of the matter traversed''. 
	This was called the {\it soft model}.
 	However ``a simple calculation based on probability shows that
	the chance of an $\alpha$ particle being deflected through 90$^{\circ}$
	is vanishingly small''. 
	Rutherford continued ``It seems reasonable to
	assume that the deflection through a large angle is due to a single
	atomic encounter, for the chance of a second encounter of the type to 
	produce a large deflection must be in most cases exceedingly small.
	A simple calculation shows that the atom must be a seat of an intense
	electric field in order to produce such a large deflection in a single
	encounter''.
	According to the
	soft model the distribution of scattered particles should fall off
	exponentially with the angle of deflection; the departure of the
	experimental distribution from this exponential form 
	is the signal for hard scattering. The soft model was wrong.
	Geiger and Marsden found that 1 in 20,000 $\alpha$ particles was
	turned at 90$^{\circ}$ or more in passing through a thin
	foil of gold; The calculation of the soft model predicted one in
	10$^{3500}$.
	 The nuclear atom was born with a hard constituent , the
	small massive positively charged nucleus. 
 	Rutherford calculated the angular distribution expected
	from his nuclear model and he obtained the famous
	$\sin^{4}(\phi/2)$ law, where $\phi$ is the scattering angle.

	At almost the same time Bohr (1913) proposed the model for the
	dynamics of the nuclear atom, based on a blend of classical mechanics
	and the early quantum theory. This gave an excellent account of the
	spectroscopy of the hydrogen. Beginning with the experiments of Frank
	and Hertz the evidence for quantized energy levels was confirmed in
	the inelastic scattering of electrons from atoms.

	The gross features of atomic structure were described well
	by the nonrelativistic quantum mechanics of point-like electrons
	interacting with each other and with a point-like nucleus, via Coulomb
	forces.

	As accelerator technology developed it became possible to get beams of
	much higher energy.  
	From the de Broglie $\lambda=h/p$ relationship 
	it  is clear that the resolving power of the
	beam becomes much finer and deviations from the Rutherford formula 
	for charged particles scattering (which assumed point-like $\alpha$'s
	and point-like nucleus) could be observed. And they did! 
	At SLAC in 1950 an electron beam of 126 MeV
	(instead of $\alpha$'s) was used on a target of gold
	and the angular distribution of the electrons 
	scattered elastically fell below the point-nucleus prediction.
	(qualitatively this is due to wave mechanical diffraction effects
	over a finite volume on the nucleus). The observed distribution 
	is a product of two factors: the scattering from a single point-like 
        target ({\it a la Rutherford} with quantum mechanical
	corrections, spin, recoil etc.) and a ``form factor'' 
	which is characteristic of the spatial extension of
	the  target's charge density.

	From the charge density distribution  
	it became clear that the nucleus has a charge radius
	of about 1-2 fm (1 fm=$10^{-15}$ m); In heavier nuclei 
	of mass number $A$
	the radius goes as $A^{1/3}$ fm. If the nucleus has a finite spatial 
	extension, it is not point-like. 
	As with atoms, inelastic electron scattering from nuclei 
	reveals that the nucleus can be excited into a sequence of	
	quantized energy states (confirmed by spectroscopy).

	Nuclei therefore must contain constituents distributed over a size of
	a few fermis, whose internal quantized motion lead to the observed
	nuclear spectra involving energy differences of order a few MeV.

	Chadwick discovered the neutron in 1932 establishing
	to a good approximation that nuclei are 
	neutrons and protons (generically called
	nucleons). Since the neutron was neutral, a new force with range of
	nuclear dimensions  was necessary to bind the nucleons in the nuclei.
	And it must be very much stronger than the electromagnetic force 
	since it has to
	counterbalance the ``uncertainty'' energy ($\approx$20 MeV; the
	repulsive electromagnetic potential energy 
	between two protons at 1 fm distance is
	ten times smaller, well below the nucleon rest energy).

	Why all this? The typical scale of size and energy are quite different
	for atoms and nuclei. The excitation energies in atoms are in general
	insufficiently large to excite the nucleus: hence the nucleus appears
	as a small inert point-like core. Only when excited by
	appropriately higher energy  beams it reveals that it has a
	structure. Or in other words the nuclear degrees of 
	freedom are frozen at the atomic scale.

%
%
	To figure out whether the nucleons are point-like 
	the picture we already developed is repeated once again.
	First elastic scattering results 
	of electrons from nucleons (Hofstadter)
	revealed that the proton has a well defined form factor,
	indicating approximately exponential 
	distribution of charge with an $rms$ radius 
	of about 0.8 fermi. Also the magnetic moments of the nucleons
	have a similar exponential spatial distribution. Inelastic scattering
	results, as expected, showed signs of nucleon spectroscopy
	which could be interpreted as internal motions of constituents.
	For example in the scattering of energetic electrons from protons 
	there is one large elastic peak (at the energy of the electron beam)
 	and other peaks that correspond to excitations of the recoiling
 	system. The interpretation of the data is a hairy story:
	several excited recoil states contribute to the same peak
	and even the apparently featureless regions conceal structure. In one 
	of the first experiments that used close to 5 GeV electrons, only 
	the first of the two peaks beyond the elastic had a somewhat simple
 	interpretation: It corresponded exactly to a long established
 	resonant state observed in pion-nucleon scattering and denoted by
 	$\Delta$. Four charge combinations correspond to the
 	accessible  pion-nucleon channels: $\pi^{+}p,~\pi^{+}n (\pi^{0}p),
	~\pi^{-}p~ (\pi^{0}n),~\pi^{-}n$. 
	 The results of such ``baryon spectroscopy'' experiments
	 revealed an elaborate and parallel scheme
	as the atomic and nuclear previously. One series of levels comes in two
	charged combinations (charged and neutral) and is built on the proton
	and neutron as ground states; the other comes in four charge
	combinations (--,0,+,++) with the $\Delta$s as ground state.

	Yukawa predicted the pion as  the quantum of the short-range nuclear
	forces. In the 60s the pion turned out to 
	be the ground state of excited states
	forming charge triplets.  Lets note again that we are 
	looking at the excitations of the 
	constituents of composite systems. 
	Gell-Mann and Zweig proposed that the nucleon-like states 
	(baryons), are made of three 1/2 spin constituents:
	the quarks. The mesons are quark-antiquark bound states. Baryons and
	mesons are called collectively hadrons. As in the nuclear case the
	simple interpretation of the hadronic charge multiplets is that the
	states are built out of two types of constituents differing by one unit
	is charge - hence its constituents appeared fractionally charged.
	the assignment was that the two constituents were the up (u) and down
	(d) with 2/3 and -1/3 charge. The series of proton would then be $uud$ 
	and the neutron $udd$ while the $\Delta^{++}$ would be $uuu$.
	The forces between the quarks must be charge independent to 
	have this kind of excited states. Lets point here that the typical 
	energy level differences in nuclei are measured in MeV. For the 
	majority of nuclear phenomena  the neutrons and protons remain in
	their ground unexcited states and the hadronic excitations are being
	typically of order MeV; the quark-ian, hadronic degrees of
	freedom are largely frozen in nuclear physics.
	People did not like the quarks; They really thought it was nonsense. 
	Nevertheless a very simple ``shell model'' approach of the nucleon was
	able to give an excellent description of the hadronic spectra
	in terms of three quarks states  and bound states of quark-antiquark.
	So now, how would we try to see
	 those quarks directly? We need to think 
	Rutherford scattering again and
	extend the inelastic electron scattering measurements to larger
	angles. The elastic peak will fall off rapidly due to the exponential
	fall off of the form factor. The same is true for the other distinct
	peaks; this indicates that the excited nucleon states have some
	finite spatial extension. The bizarre thing is that for large energy
	transfer the curve does not fall as the angle increases.
	In other words for large enough energy transfer the electrons
	bounce backwards just as the $\alpha$s did in the Geiger-Marsden 
	experiment.  This suggests {\it hard} constituents! 
	This basic idea was applied in 1968 
	in so-called {\it deep inelastic scattering} 
	experiments by Friedman, Kendall and Taylor 
	in which very energetic electrons were scattered off of protons. 
	The energy was sufficient to probe distances shorter 
	than the radius of the proton, and it was discovered that 
	all the mass and charge of the proton was concentrated 
	in smaller components, spin 1/2 hadronic constituents
	called ``partons'' which were later identified with quarks. 
	A lot of spectroscopy and scattering data became 
	available in the '70s
	and still people used the quarks as mathematical elements that 
	help systematize a bunch of complicated data. 
	In fact the following quote is
	attributed to Gell-Mann: ``A search for stable quarks of charge -1/3 or
	+2/3 and/or stable di-quarks of charge -2/3 or +1/3 at the highest 
	energy accelerators would 
	help reassure us of the non-existence of real quarks''.
	Indeed quarks have not been seen as single isolated particles.
	When you smash hadrons at high energies, where you 
	expect a quark what you observe downstream is a lot more hadrons
	- not fractionally charged quarks. 
	The explanation of this quarky behavior -- that they don't exist 
	as single isolated particles but only as groups 
	``confined'' to hadronic volumes --
	lies in the nature of the  interquark force.
 	November 1974 was a revolution of quarks: A new series of mesonic
	(quark-antiquark) spectra,  the $J/\psi$ particles, were
	discovered, with quantum number characteristics  of
	fermion-antifermion states.   The $J/\psi$ spectrum is very 
	well described in terms of $c\bar c$ states where $c$ 
	is a new quark:  the
	{\it charm}. The $c\bar c$ is called {\em charmonium} 
	after the $e^{+}e^{-}$
	{\em positronium}. There is a funny resemblance between the energy 
	states of the charmonium with those of the positronium given that 		the positronium is bound via electromagnetic forces while the
	charmonium via strong forces.

	Back to Rutherford again. We saw how the large angle large energy
	transfer electron scattering from nucleons provided evidence for
	``hard constituents''. What if we collide two nucleons?
 	With a ``soft'' model 
	of the nucleons we expect some sort of an exponential fall
	off of the observed ``reaction products'' 
	as a function of their angle to 
	the beam direction. On a ``hard'' model we should see prominent
	``events''  at wide angles, corresponding to collisions between
	the constituents.
	The hard scattered quarks are
	converted into two roughly collimated {\it 'jets' of hadrons}.
	These jets and their angular distributions provide indirect evidence
	of quarks. At $p\bar p$ collisions at CERN  jets were observed 
	in the 80s when CERN achieved with SPS the largest momentum transfers.
	Clear evidence of hadronic jets  associated with primary quark
	processes  were observed earlier in electron-positron collisions.

	If quarks are not point-like we expect to see at  higher
	energies, where the sub-quarkian degrees of freedom unfreeze
	perhaps at the Tevatron and the LHC,  deviations
	from the theory similar to the deviations
	observed in the deep inelastic scattering experiments.

\subsection{Technical handbook}

The fragments of a high energy collision in matters of nanoseconds
have decayed and/or left the detectors. In the early times of 
particle experiments an event was a picture in 
a bubble chamber for example,
of the trail the particles left when ionizing a medium.
Now an event is an electronic  collection of the 
trails many particles left 
in a multiple complex of detectors. 
We are taking a little detour here to 
define some  jargon particle physicists use, to 
discuss briefly what is the modern process of particle detection
and what is the data analysis after all. 
To start, there is
a physics collision for which we have a theoretical model to describe 
the particle interaction (we draw a Feynman graph). The fragments 
of the collision decay and interact with the detector 
material (we have for example multiple scattering).
The detector is responding 
(there is noise, cross-talk, resolution, response function, alignment,
temperature, efficiency...) 
and the real-time data selection (trigger) 
together with the  data acquisition system give out 
the raw data (in the form of bytes;
 we read out addresses, ADC and TDC values and bit patterns).
 The analysis consists of converting the 
raw data to physics quantities.
 We apply the detector response (e.g. calibration, alignment)
and from the interaction with the detector material we perform pattern 
recognition and identification.
From this we reconstruct the particles' decays 
and get results based on 
the characterization of the physics collision. 
We compare the results with 
the expectations by means of a reverse path that simulates the physics
 process calculated theoretically and driven through a computational model
 of the detector, the trigger and data acquisition path (Monte Carlo).  
The challenge is to select the useful data and record them with 
minimum loss (deadtime) when the detector and accelerator is running properly;
and of course to analyze them and acquire results that
are statistically rather than systematically limited. When 
the statistical uncertainty becomes smaller than
the systematic uncertainty, it is time to build a new
experiment for this measurement.

To give an example, in a hadronic collision 
at CMS in 2007, the interaction rate 
is 40 MHz (corresponding to data volume of 1000 TB/sec). 
The first level of
data selection  is hardware implemented  and 
by using specific low level analysis  
reduces the data to  75 kHz rate 
(corresponding to data volume of 75 GB/sec). 
The second level of 
judging whether an event is going to be further retained is implemented 
with embedded processors that 
reduce the data rate to 5 Hz (5 GB/sec). 
The third level of the trigger is a 
farm of commodity CPUs that 
records data at 100 Hz (100 MB/sec). 
These data are being recorded for offline analysis.
The final data volume depends 
on the physics selection trigger and for example 
we expect at the 
first phase of the Tevatron Run II to have between 1 and 8 petabytes 
per experiment.
The improvement in high energy experiments is multifold. 
Better accelerator design and controls
give higher energies and collision rates.
 Better trigger architecture 
is making best use of the detector subsystems. Better storage, 
networks and analysis 
algorithms contribute to precise and 
statistically significant results.
 Large CPU and clever algorithms 
improve the simulations and theoretical 
calculations that in turn enable the 
discovery of new physics in the data. 

To summarize the jargon: {\it trigger} is a fast, rigid and primitive, 
usually hardware implemented selection applied on raw data or
even analog signals from the detectors. 
Triggers have {\it levels} that an event passes
through or fails. The trigger system and algorithms are 
 emulated so that the
signals are driven through a computational model of the 
trigger path, and the results are compared with
the acquired data for diagnostic purposes.
The efficiency of a multilevel trigger path
is measured in datasets coming from orthogonal
trigger paths (e.g. you want to measure 
the efficiency of a multilevel missing energy 
trigger in a dataset of events that come
from a trigger that has no missing energy
in its requirements).
The duration of time when the data acquisition 
system cannot accept new data
(usually because it is busy with current data) is called 
{\it deadtime}. To go faster 
modern experiments are running parallel data acquisition on 
sub-detector systems that are ending in 
fanned out triggers ({\it data streams});  
the combination of the fragments of data from the 
detector subsystems is the {\it event building}. 
A {\it filter} is a later selection after the trigger,
which is usually software implemented and can be sophisticated.  
{\it Reconstruction} is
the coding that converts the 
sparsified hardware bytes to physics objects (tracks, vertices,
tags etc.) A computer {\it farm} is a dedicated 
set of processors and associated 
networks used to run filters, event reconstruction, 
simulation etc. {\it Efficiency} is the 
probability to pass an event (signal or background). {\it Enhancement} 
is the enrichment of the data sample after a selection is applied.

\cleardoublepage
\section{Bottom-top experimental approach} 

	There is a lot of unfinished business 
	before we 
	confront strings and extra dimensions
	in colliders. However, we investigate 
	phenomena in fundamental physics 
	 not diagram by diagram, but scale by scale 
	as it has been 
	pointed out by Joe Polchinski in his string colloquium 
	~\cite{polstring}. We do this in theory and unavoidably in 
	experiment too: a fact that is punctuating the
	role of higher energy accelerators.  
	In the section discussing the	
	path from atoms to quarks, we saw that 
	different degrees of freedom are frozen 
	at different energy regimes and unveiled in others.
	 We also saw that the energy scale
	at which these degrees of freedom show up 
	is only discovered by means of experimental data.
	It is very difficult to theoretically determine 
	the energy scale where new phenomena 
	emerge (in the words of theorist Joe Lykken ~\cite{aaas_jl}),
	 in particular when one 
	has no idea what these phenomena are, and even if one 
	has a good scenario of what they may be.
	In the past few years space 
	itself is approached as degrees of freedom 
	that may unfreeze and emerge
	at a particular characteristic energy scale.

	In particle physics we talk about two major scales, 
	one of which we have extensively studied theoretically 
	and experimentally. It is the electroweak 
	scale (10$^{-17}$ cm or 10$^{11}$ eV)  where 
	three of the four forces of 
	nature have comparable strengths and the masses of the W and Z
	bosons are generated. 	The other distinct scale
	 is the Planck scale which is very distant in value from
	the electroweak scale (10$^{-33}$ cm or 10$^{27}$ eV) 
	and is the scale at which gravity would have 
	comparable strength with the 
	rest of the known forces of nature. 
	At the electroweak scale all physical phenomena 
	except for gravity 
	are very well described by the Standard Model 
	of the elementary particles and their interactions.
	At the scale of 100$\mu$m gravity follows 
	Newton's law \cite{JHMS:02} and at large length scales 
	the general theory of relativity takes over.
	 With the present data and understanding 
	of physics at the electroweak scale we can 
	 argue that a compelling completion
	of the standard model is found in supersymmetry, 
	discovered by Pierre Ramond \cite{pierre:71} Wess, Zumino, 
	and others
	in the seventies as a side-effect of theoretical 
	attempts to include fermions in string theories.
	Supersymmetry is a good theory that gives back 
	more than the inputs it requires. The theory is 
	supremely decorated, among others with the best candidate 
	for the dark matter of the universe, the path towards 
	the inclusion of gravity in a unified way 
	with the rest of the forces, the exquisite prediction of 
	the GUT scale where all couplings meet, 
	the accurate and  precise 
	(within 1$\%$ of its measured value) 
	predicted value of the weak angle 
	at the electroweak scale from the GUT scale, 
	the heaviness of the top quark and its function
	in electroweak symmetry breaking through
	 radiative corrections, and even clues towards
	the understanding of the negative pressure
	 that drives the recently discovered accelerating 
	universe. We must keep 
	in mind the fact that all the above can 
	be realized with supersymmetry at the electroweak 
	scale, which is within the reach of 
	experiment.
	
	A variety of supersymmetric models have been 
	developed  and evolved in the past decade \cite{gordy:00}. The differences
	lying in the assumed SUSY breaking mechanism and the identity of the 
	lightest supersymmetric particle. Models with 		
	photons and leptons in the final state are usually
	 gauge mediated, models with, 
	``disappearing'' tracks are signatures 
	of anomaly mediation while  jets,  leptons and missing energy
	are signatures of generic minimal supersymmetric models. 
	Although supersymmetry is the best bet for the 
	next great scheme that will include and further the 
	Standard Model description of nature, 
	there is hardly a good 	
	realizable model, {\it the} model, to be put to the test. 
	Clearly this will be accomplished when we discover 
	the first supersymmetric 
	particles. However it is of critical importance to
	 use the data we have in the same way the 	
	blind use their walking stick to go about. 
	Let me give an example of what I mean. In the 
	time when LEP results were pouring in, 
	the visionaries expected supersymmetric particles right 
	around the corner. Corner by corner there was no sign 
	of them. Within the supersymmetric models
	developed at that time,  and using the data acquired, 
	they discovered false theoretical 
	assumptions and corrected them.
	With the new assumptions in the models the vision
	 was moved to a different value of the similar 
	scale. Now we expect supersymmetric particles to
	 be found at the Tevatron and the LHC. In other words there 
	is an almost divine feedback mechanism between data,
	 their studious interpretation and the tracking 
	of the theory. The data is sculpting the theory. 
	Experimentalists are comparing the observed data
	with the standard model predictions and use the result
	 of the comparison to 
	probe particular theoretical models. 
	The colorful complicated exclusion and reach plots  
	that are generated, although  attempted to be as 
	model independent as possible, are always
	increasing our understanding of a particular theory 
	and why it might not be the proper one.
	If supersymmetry- despite being the most educated 
	formulation of the biggest theory we are looking 
	for- is wrong, the data will show it and will point
	 to a better theory. Rather curious as it may seem at first look
	I completely agree with E. Witten that  
	``. . . One of the biggest adventures of all is the search for 
	supersymmetry''\cite{gordy:00}.

	The bottom-top experimental approach in the title of this section 
	refers to the upwards evolving  energy
	of the machines we build and correspondingly the 
	energy scale we explore.
	However nature is tricky, and so for example 
	we have measured the mass of the heaviest
	of the quarks before the masses the tiniest 
	of the leptons, the neutrinos.
	There is always a chance that although machine-wise we work 
	our way up in energy, the physics of the highest 
	energies shows up unexpectedly.
	This is the case in the scenarios of large 
	extra dimensions that have been
	extensively studied in this summer school, 
	where the real Planck scale is indeed
	allowed to take values close to the electroweak scale. 
	The same is true
	for the physics of strings and black holes in high energy collisions.

	In the next section I will give an example of a 
	a data analysis 
	that was interpreted in two SUSY models, 
	one more and one less constrained, and the ingredients
	that enter this analysis. And similarly for 
	an extra dimensions analysis.
	I would also like to refer you to a marvelous report on particle 
	physics for theorists that appeared in 
	1996 TASI proceeding by Persis Drell \cite{persis:97}.

\subsection{Example 1: From a trigger, to a model, to a signature,
 to the data, and back to a model}

	The ``missing energy plus jets'' signature is 
	referred to as one of the golden
	signatures when searching for SUSY in hadron colliders. 
	The reason is the large rate
	at which squarks and gluinos are produced
	and the abundance of the lightest supersymmetric particle
	in the decay chain of supersymmetric particles. 
	The large missing energy would originate from the two LSPs 
	in the final states of the squark and gluino 
	decays. The three or more  hadronic jets  would result 
	from the hadronic decays of the $\sq$ and/or $\gls$. 
	It is not only SUSY that 
	gives you such a signature. 
	Leptoquark or technicolor models can have the same or 
	similar final state.
	The general analysis direction for the search is 
	similar for different models and
	the data reduction steps are kept as inclusive as possible. 
	However, when optimizing an analysis
	a model is chosen, in this case  supersymmetry.
	
	We do use missing energy collider data  
	for measurements of the $Z$ boson 
	invisible decay rates ~\cite{markel}, 
	the top quark cross section \cite{top_xmet}, 
	searches for the Higgs boson ~\cite{kruse} 
	and other non-\sm~ physics processes
	 ~\cite{stopsb}, ~\cite{castro},~\cite{lq}.

	In a detector with hermetic  4$\pi$ solid angle coverage the
	measurement of missing energy is  the measurement 
	of neutrino energy plus the 
	energy of any unknown weakly interacting particle..  
	In a real detector it is also a measurement 
	of energy that escapes detection
	due to uninstrumented regions.  Jets dominate
	the cross section for high \pt~ proton-antiproton 
	scattering and contribute 
	to a final all-hadronic state with missing energy  
	from the 
	heavy flavor decays but mostly from jet energy 
	resolution and mismeasurements.
	QCD  production  (including \ttb) and $W/Z$ QCD 
	associated production dominate in a sample
	of events with large missing energy and multiple jets. 
	This is because of the 
	invisible $Z$ decays and the decays of the 
	$W$ to a charged lepton and a neutrino. 
	These processes can be understood with the  data
	before narrowing the search for any exotic 
	signal with the same
	detection signature.  

	In this particular example  the  {\it blind box} 
	method was employed. 	
	It was discussed by  R. Cousins and others 
	in the past decade and various 
	versions and improvisations of the method have been used in different 
	measurements and searches.

	Important data analysis methods to identify in this example are 
	a) all data samples are revertexed and reclustered offline	
	according to the best determined hard scattering vertex; 
	b) on all  Monte Carlo samples a tracking degradation 
	algorithm is applied to appropriately account for the tracker
	aging c) a two stage cleanup algorithm is applied to 
	reduce a sample that has a two-to-one
	noise to physics ratio  and make it useful for further analyses; 
	d) orthogonal data samples are used to normalize the 
	theory predictions for most of the Standard Model 
	backgrounds. In all predictions
	the appropriate trigger efficiencies 
	measured in the data are folded in before the 
	comparison with the data and extraction of the 
	normalization factor. Notice that when you normalize to the 
	data the uncertainty on the normalized predictions 
	comes mostly from the statistics of the
	data samples used,  e) in the case of a blind box analysis,  
	we form control regions around the box 
	by inverting the requirements which define it and compare  
	the data in these regions  
	with the normalized prediction to make sure 
	that the normalization of the background samples 
	are accurate and to diagnose potential pathologies 
	before opening the box. The same
	is true if the analysis in not blind: 
	control regions should be checked for the validity of
	the standard model predictions. 
	The main objective of a blind analysis is to 
	avoid biased human decisions involving 
	the data selection. We achieve this by insulating 
	the signal candidate data sample until 
	we estimate the total background. 
	We then {\it a priori} define the signal 
	candidate data sample based on the signal 
	signature and the total  background 
	estimate and precision. In this analysis we use 
	three variables to define 
	the signal candidate region : The missing transverse energy, 
	\met, the scalar sum $\Ht \equiv E_{T(2)}+E_{T(3)}+\met$, 
	and the isolated track
	multiplicity, \niso ~\cite{niso}.  
	The blind box contains events
	with $\met\ge 70$ GeV, \mbox{$H_T\ge 150$ GeV}, 
	and \niso=0.  Several questions arise
	pertaining to the choice of the variables and their values.
	The value 70 GeV is chosen based on the MET trigger efficiency. 
	We need to know
	how many standard model background events and how 
	many for example SUSY events
	would pass through the same triggering scheme 
	and be registered in the data 
	sample that we are examining, in this case the \met\ sample. 
	We normalize the standard model backgrounds using 
	differently triggered samples (e.g lepton triggered samples),
	 we apply the 	normalization to the prediction and we 
	also fold in the \met\ trigger efficiency.
	More than 95\% of  events with reconstructed \met\ above 70 GeV
	would have passed the \met\ trigger during data taking. 
	We extract this 
	from jet data samples that do not 
	require missing energy in their trigger path.
	Is this the optimal value for searching for new physics?  
	Starting with
	any value less than 70 GeV would introduce larger uncertainties from
	the trigger efficiency. 
	For values above 70 GeV we need to optimize the signal 
	to background ratio for a particular signal. 
	The \Ht\ variable is 
	constructed so as to play a discriminant role between the signal
	and the standard model backgrounds. 
	Notice that the sum does not include the 
	first jet which is the well measured jet. 
	This is so that we discriminate
	events with real missing energy from events where the missing energy 
	is a result of jet (second and third usually) mismeasurements. At LHC
	just the sum of all jets in the event would be a good variable.
	 The track isolation	
	variable is a counter of the number 
	of high $p_T$ isolated tracks. By requiring 
	it be zero we  indirectly veto event with leptons, 
	this way focusing the search to all-hadronic states, and 
	f) for processes where the measured or theoretical cross sections 
	have large uncertainties such as most QCD processes 
	we need to find adequate ``standard(izable) 
	candles''; a good example is the dielectron+jets 
	event data sample where the invariant mass of the 
        pair is consistent with the Z boson mass.

        The parts of the analysis are: 
	\begin{itemize}
        \item The data Pre-Selection, designed to  acquire  a 
	high purity sample of large real missing energy data (this is
 the cleanup from the junk that end up in the sample: 
everything that goes wrong in 
the detector as well as cosmics, end up in this sample). 
The result of this cleaning 
	of the sample is shown in Figure ~\ref{fig1}.
\begin{figure}
\centerline{
\psfig{figure=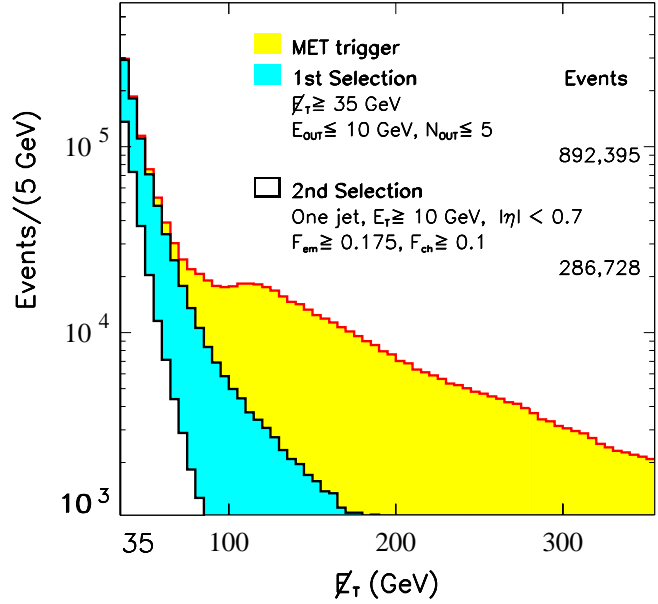,width=3.7in}}
\caption{ The \met~spectrum after the online trigger
and the two stages of the data preselection. The numbers 
of events surviving the
first and second selections are 892,395 and 286,728, respectively. The
variables $E_{OUT},~N_{OUT}$ are energy and number of towers 
out of time~\protect{\cite{smaria_thesis}}.}
\label{fig1}
\end{figure}
	\item The $W$ and $Z$ boson QCD associated production background 
estimate. As mentioned,
	the  $Z(\rightarrow e^{+}e^{-})+jets$ data sample is used as standardizable candle to normalize 
	the theoretical rate predictions. The sample used is the events selected from a high
	energy electron trigger that have a second electron and the invariant mass of the 
	the two is consistent with the Z mass as shown in Figure ~\ref{fig2}. Exactly the same
	selection rules and corrections are applied in the data and Monte Carlo samples.   
\begin{figure}
\centerline{
\psfig{figure=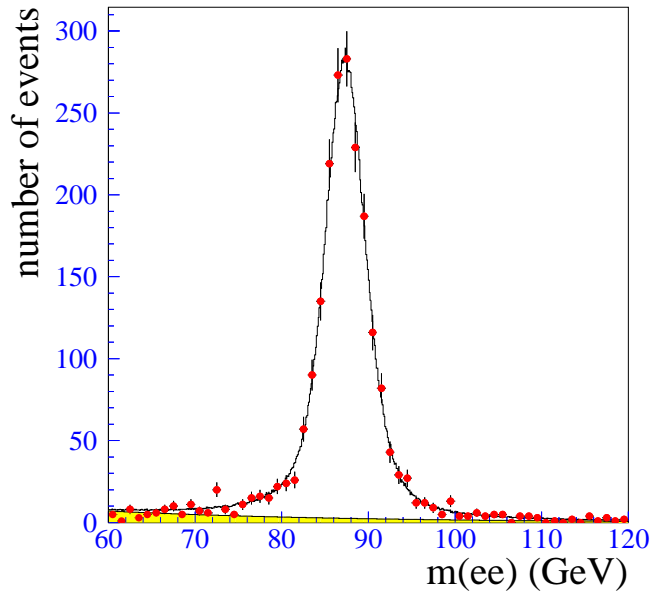,width=3.4in}}
\caption{The $Z^0$ mass as reconstructed in the mode
$Z^0\rightarrow e^+e^-$ by the D\O\ detector.  The
shaded region at the bottom of the plot is the background
contribution.  The peak does not fall exactly on the
true value of $M_Z$ because not all of the energy
corrections have been applied to the data. This is why it is referred to in the text as a {\it standardizable candle}; After the appropriate corrections are made the
events in the Z boson mass peak are used  for the normalization 
of the the vector boson+jets rates. The histogram is the prediction and the points the D\O\ data.}
\label{fig2}
\end{figure}
	\item The multijet QCD production background estimate. The JET data samples
 	are used to normalize the theoretical rate prediction. 
	The source of missing energy in QCD jet production  is the small fraction
	of $b \bar b$ and $c \bar c$ content (with the $b$ and $c$ quarks decaying
	 semileptonically) and largely the jet mismeasurements and
	detector resolution.  Analyses that require a measurement of the
	QCD multijet background use the jet data when there are extra 
requirements 
	such as a $b$-tagged jet or a lepton.  
	The data give a reliable estimate for the QCD background in
	such analyses.  Comparisons between data and \herw\ QCD Monte Carlo for the 
	$\Sigma \et$ cross section measurement at CDF   
	indicate agreement between the data and the Monte Carlo predictions.  
	In the case of the large missing energy plus multijet ($\ge 3$ jets) search
	the estimate of the QCD background is nontrivial.
	The missing energy trigger accepts QCD multijet events (45\% of the 
	online missing energy trigger are volunteers from jet triggers) 
	and the trigger threshold is too high\footnote{For Run II the 
	MET trigger now taking data, is designed with  a lower threshold (25 GeV).} to 
	allow use of the low missing energy triggered data for extraction of the
	high missing energy spectrum.  
	The high energy threshold jet triggered data with small 
	or no prescale (JET70, JET100) are not suitable
	to extract the QCD contribution to the high \met~ tails
	as they themselves constitute signal candidate samples.
	The lower energy threshold jet triggered data with large 
	prescales (JET20, JET50) are used in this analysis to estimate
	the QCD jet production contribution to the high missing energy 
	spectrum. 
	Large statistics
	3-jet QCD Monte Carlo samples are generated to simulate the JET20 and JET50 data 
	samples and used to compare the shapes of the missing energy  and
	 the $N$-jet distribution with the data.  The predictions are absolutely
	normalized to the data. 
	\item  The comparisons of the total background estimates with the data in the control regions around 
	the Blind Box. 

There are seven control regions around the blind box formed by inverting the
requirements which define it.  
We compare the \sm\ background predictions in the control regions
with the data. The results are shown in Table
~\ref{tab3}. 

\begin{table}
\begin{center}
\begin{tabular}{c|c|c|c|c}
Region Definition&EWK&QCD&All&Data\\ \hline 
{\met~$\ge$70,\Ht$\ge$150,\niso$>0$}&{14}&{6.3}&{20$\pm$5}&{10}\\ \hline
{\met~$\ge$70,\Ht$<$150,\niso$=0$}&{2.3}&{6.3}&{8.6$\pm$4.5}&{12}\\ \hline
{35$<$\met~$<$70,\Ht$>$150,\niso$=0$}&{1.95}&{135}&{137$\pm$28}&{134}\\ \hline
{\met~$>$70,\Ht$<$150,\niso$>0$}&{1.7}&{$<$0.1}&{1.7$\pm$0.3}&{2}\\ \hline
{35$<$\met~$<$70,\Ht$>$150,\niso$>0$}&{14}&{9.4}&{23$\pm$6}&{24}\\ \hline
{35$<$\met~$<$70,\Ht$<$150,\niso$=0$}&{5}&{413}&{418$\pm$69}&{410}\\ \hline
{35$<$\met~$<$70,\Ht$<$150,\niso$>0$}&{3.3}&{28}&{31$\pm$10}&{35}\\ \hline
\multicolumn{1}{c|}{Signal Candidate Region} &  & & & \\ \hline 
{\met~$\ge$70,\Ht$\ge$150,\niso$=0$}&{35}&{41}&{76$\pm$13}&{74}\\ 
\end{tabular}
\caption{Comparison of the ~\sm~prediction and the data in the
control regions and the signal candidate region (blind box). 
After the contents of the control regions 
were compared in detail to standard model predictions, 
we opened the box and found 74 events. (\met~ and \Ht~ in GeV.)}  	
\label{tab3}
\end{center}
\end{table}

\end{itemize}
Of the 76 events predicted in the blind box, 41 come from QCD 
and 35 from electroweak processes. Of the latter we estimate
$\sim$37\% coming from
$Z(\rightarrow \nu\bar{\nu})+\ge 3$ jets, $\sim$20\% from
$W(\rightarrow \tau\nu)+\ge 2$ jets, $\sim$20\% from the combined
$W(\rightarrow e(\mu)\nu_{e}(\nu_{\mu}))+\ge 3$ jets, and $\sim$20\%
from \ttb\ production and decays. We also compare the kinematic
properties between \sm\ predictions and the data around the box and 
find them to be in agreement ~\cite{smaria_thesis}.

	As we mentioned in the introduction we can stop 
	here and show histograms
	of how well the data match the Standard Model predictions
	both in the blind box and in the control regions \ref{fig2}.
\begin{figure}
\label{fig2}
\centerline{
\psfig{figure=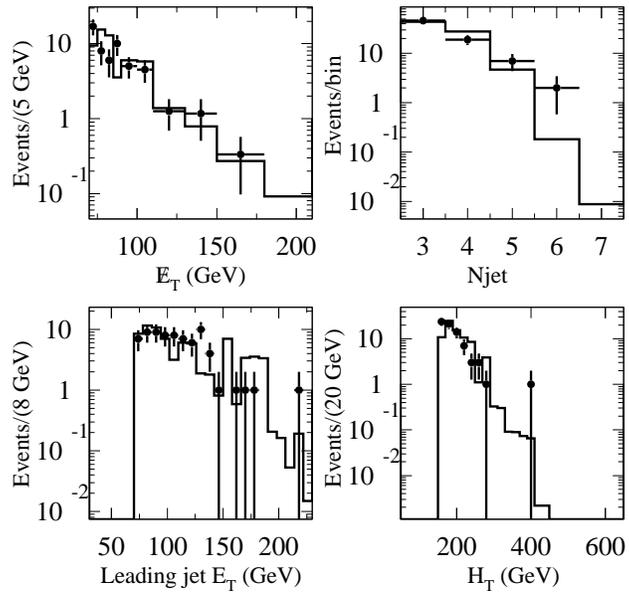,width=3.7in}}
\caption{ Comparison in the blind box
between data (points) and \sm\ predictions (histogram) of ~\met,
$N_{jet}$, leading jet \et\ and \Ht\ distributions. There are 74
events in each of these plots, to be compared with 76$\pm$13 SM
predicted events. Note that the \met\ distribution is plotted with
a variable bin size; the bin contents are normalized as labelled.}
\end{figure}
	
	We can combine
	the level of agreement in this channel  
	with a number of searches in  other final 
	states and make a global analysis of
	how well all the data in all analyses  
	match the Standard Model 
	predictions. However we choose to take the analysis 
	a step further. 
	We pick two models and provide an answer to a model builder
	who wants to know how light a gluino is allowed to 
	be based on this particular search.
	We study the hadroproduction of scalar 
	quarks and gluinos and all their decays in
        minimal Supersymmetry and Supergravity frameworks. 
	Important theoretical considerations 
	that need to be underlined are a) the use  of $\tan\beta=3$ 
	to generate datasets of squark and gluino events, b) the
	study of the production  of only thefirst two 
	heavy generations of 
	squarks ($\tilde{u},\tilde{d},\tilde{c},\tilde{s})$
	 in the general MSSM
	framework while in mSUGRA  the production of the bottom squark
	($\tilde{b}$) is also considered. Work which is now underway shows that
	the results are valid also for $\tan\beta=30$.

Based on the observations, the \sm\ estimates and their uncertainties,
and the relative total systematic uncertainty on the signal
efficiency, we derive the 95\% C.L. upper limit on the
number of signal events. The bound is shown on the $\msq-\mgls$ plane in
Figure 3.  For the signal points generated with mSUGRA, the limit is
also interpreted in the $M_{0}-M_{1/2}$ plane ~\cite{smaria_thesis}.

\begin{figure}
\label{fig3}
\centerline{
\psfig{figure=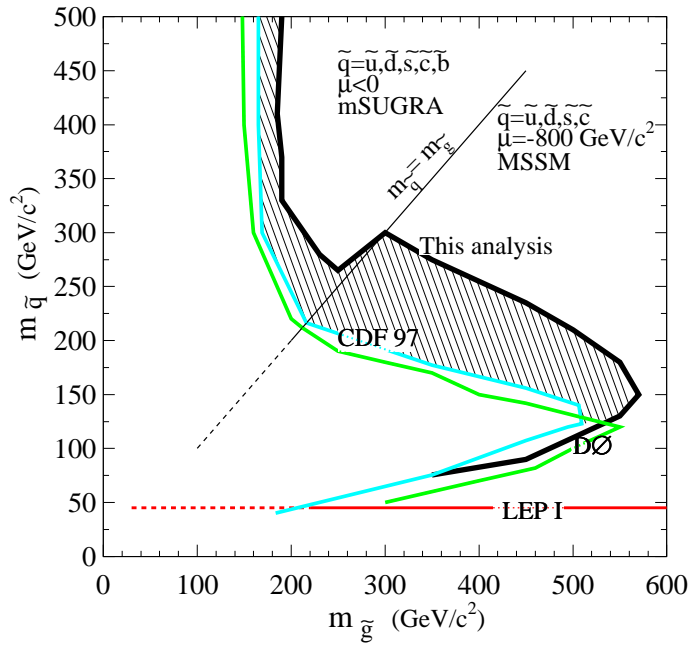,width=3.9in}}
\caption{The 95\% C.L. limit curve in the $\msq~-~\mgls$ plane
for $\tan\beta=3$; the hatched area is newly excluded by this
analysis. Results from some previous searches are also
shown.}
\end{figure}
 
\subsubsection{What did we learn?}

We learned that there is no excess of multijet events at the
tails of the missing energy distribution in 84 pb$^{-1}$ of data.
We learned that we can simulate well the tails of the missing energy
adequately for 84 pb$^{-1}$ of data and predict the standard model
backgrounds robustly. We learn that the LEP results are in
accord with the Tevatron results when we both consider a minimal 
supersymmetric model. And we learn that 
the scale of SUSY can still be the electroweak
scale. Lets discuss  the latter  a bit more. 

The notion of naturalness and fine-tuning ~\cite{wilsonetc} 
is not extremely  well defined, but it is used  unavoidably since
it is the fine-tuning in the \sm~ that motivates
low-energy supersymmetry and supports the projection that
superpartners should be found before or at the LHC.
Many measures and studies of fine-tuning  have appeared
in the literature ~\cite{barbguid,savas_2,anderson,gordy,feng_machev}.

In a model-independent analysis ~\cite{Coleman:1973sx}, naturalness constraints are weak for
 some superpartners, {\it e.g.} the squarks and sleptons of
the first two generations. In widely considered scenarios with
scalar mass unification at a high scale, such as minimal supergravity, it is
assumed that the squark and slepton masses must be $\lappeq$1 TeV/$c^2$.
This bound places all scalar superpartners within the reach of present and near
future colliders. This assumption is re-examined ~\cite{feng_machev}  in
models with strong unification constraints, and the squarks and sleptons
are found to be natural even with masses  above 1 TeV/$c^2$.
Furthermore by  relaxing the universality constraints the
naturalness upper limits on supersymmetric particles increase
significantly ~\cite{savas_2} without extreme fine-tuning.
This suppresses sparticle mediated rare processes and
the problem of SUSY flavor violations is ameliorated. The fine-tuning due to
the chargino mass is found to be model dependent ~\cite{gordy}.
With or without universality constraints the gluino remains
below 400 GeV/$c^2$ ~\cite{savas_2}. In fact, as it is pointed in ~\cite{gordy}
the tightest constraints on fine-tuning come from the
experimental limits on the lightest CP-even Higgs boson and the
gluino for a number of supersymmetry models. It is then
those two key particles that are within reach
of the Tevatron collider.
 These results
follow from the observation that fine-tuning is mainly dominated by
$M_3$, the gluino mass parameter at the electroweak scale,  and this dominant contribution can be partly canceled
by negative contributions from other soft parameters, as can be seen
from the expansion of the $Z$  mass in terms of the input parameters
(and for fixed $\tan\beta=2.5$) \cite{gordy} :

$$
M_{Z}^{2}=-1.7\mu^{2}(0) ~+~7.2M_{3}^{2}(0)~-0.24M_{2}^{2}(0)~ +~ 0.014M_{1}^{2}(0) + ...
$$

The required cancellation is easier if $M_{3}(0)(\sim \mgls)$
is not large (or alternatively  if $M_2$ is increased for a given $M_3$).
Using the results  of this analysis on the gluino mass we get: 

$$M_3\gappeq 300 \rightarrow  \frac{7.2M_{3}^{2}}{M_{Z}^{2}} \gappeq 80$$

A similar relation is derived using the LEP limits on the
chargino $m_{\stilde{\chi}_{1}^{+}}\gappeq 100$ GeV/$c^2$  ~\cite{LEPC2000,martinDPF,kane:02} that points to the consideration of gaugino mass non-unification
with a lighter gluino. The main effect of a relatively light gluino
is the enhancement of the missing energy plus multijet signal
with a lepton veto, since for a given chargino mass not yet excluded, the
$\gls\gls$ cross section is enhanced and the gluino cascade decays through charginos  are suppressed (fewer leptons are produced in the final state
\cite{sug_rep,Anderson:2000ui}).

\subsubsection{The \met~ trigger}
The \met\ trigger drives a number
of analyses a few of which 
are the following:
\begin{itemize}
{\item {\bf Vector boson} production and leptonic decays.
Although there is a  dedicated
\met\ plus lepton trigger for the study of the $W$ boson,
$W$  QCD associated production remains a crucial
background for a number of  searches beyond the \sm\ and the
\met\  plus jets trigger provides a good sample to study
these processes.
 For $Z$ production and decay,
the \met\ sample provides a dataset to measure directly the
$Z \to \nu\bar\nu$ + jets cross section.
Furthermore again,  the $Z$ boson QCD associated production is a
background to many  searches.}
{\item {\bf top quark}  production and decay to a $W+b$. The \met\
trigger provides an alternate dataset to measure the top cross section.}
{\item {\bf Associated Higgs-$W$ and Higgs-$Z$ production}. The \met\
combined with a b quark tagging or a tau lepton tagging trigger
can provide a highly
efficient triggering scheme for the discovery of the Higgs boson.}
{\item {\bf Beyond the \sm\  searches}. To mention a few, the \met\ trigger can be used to search for:
\begin{itemize}
\item {\bf Supersymmetric partners}: In R-Parity conserving supersymmetric scenarios the LSP (Lightest Supersymmetric Particle) escapes the detector and appears as energy imbalance.
Examples are squark and gluino searches, scalar top and scalar bottom quark (utilizing an additional heavy flavor tag) searches. In gauge mediated supersymmetry breaking scenarios that incorporate gravity
the  LSP is the gravitino -- the  spin 3/2  partner of the
graviton. The gravitino goes undetected and produces energy imbalance.
\item  {\bf Leptoquarks}: The leptoquark decays to
a quark and a neutrino resulting in large \met.
\item CHArged Massive  Particles ({\bf CHAMPS}): These are long-lived
massive particles that if they are penetrating enough can go undetected  and cause
energy imbalance.
\item {\bf Gravitons}: In braneworld  theories of extra dimensions ~\cite{gn:14,ar:98}  the graviton can be produced in high energy hadron collisions and escapes to the extra spatial dimensions resulting in energy imbalance.
\end{itemize}}
\end{itemize}
The specifications  of the \met\ trigger  acquiring RUNII data was determined
using  the Run 1B data and the main feature of the trigger is the lower Level 2 \met\ threshold (25 GeV, compared to 35 GeV in Run 1B).

\subsection{Example 2: Monojets and missing energy: Looking for KK graviton emission}

By now you know exactly the steps of how to do this analysis: you need
the clean \met\ triggered data sample and orthogonal data samples to
normalize your background estimates, so that your uncertainties 
in the standard model predictions are dominated by the data statistics and
not by systematics. 

\subsubsection{On the theory}
\label{chap:intro}
It has been discussed since 1996 ~\cite{lykken:96} 
that weak scale superstrings have experimental consequences 
in collider physics. The missing energy as a signature 
was discussed already in ~\cite{ignat:98}.
Consider the braneworld  scenario where gravity propagates in the 
$4+n$ dimensional bulk of spacetime, while the rest of the standard model 
fields are confined to the 3+1 dimensional  brane.
Assume compactification of the extra 
$n$ dimensions on a torus with a common scale $R$, and identify the massive 
Kaluza-Klein (KK) states in the four-dimensional spacetime.
In such a model, the Planck scale $M_{Pl}$, the compactification scale $R$,
and the new effective Planck scale $\md$, are related by the expression:
$$        M_{Pl}^2 \sim R^n M^{2+n}_{D}      $$
where $n$ is the number of extra dimensions.

Searches for extra dimensions at colliders focus on the search for
the emission of gravitons or the effects of the exchange of virtual
gravitons (see J. Hewett, this school proceedings).  Here we focus on a  
search for graviton emission.

\begin{figure}
\centerline{
  \psfig{file=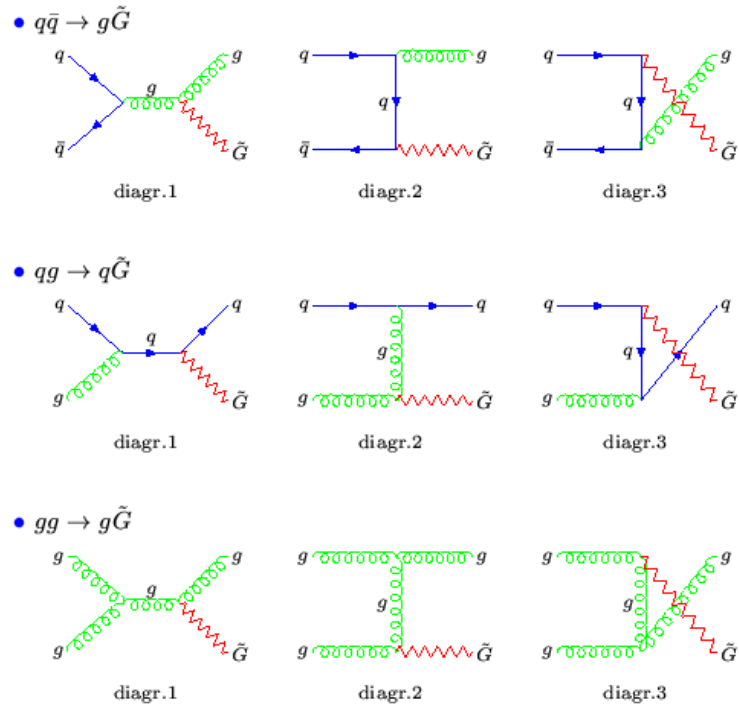,width=4in}}
   \caption{Feynman diagrams for the emission of real gravitons in
           $p\overline{p}$ collisions.
           Top: $q\overline{q} \rightarrow gG$.
           Middle: $qg \rightarrow qG$.
           Bottom: $gg \rightarrow gG$.}
  \label{feynman}
\end{figure}

There are three processes which can result in the emission of gravitons:
\begin{itemize}
  \item[]$q\overline{q} \rightarrow gG$
  \item[]$qg \rightarrow qG$ 
  \item[]$gg \rightarrow gG$ 
\end{itemize}
where $G$ is the graviton.  The Feynman diagrams for these processes are
shown in Figure~\ref{feynman}.

In all three cases the signature will be jets + $\met$.
We included in \pyt\ these processes  using the 
large toroidal extra dimensions model
of Arkani-Hamed, Dimopoulos, and Dvali \cite{ar:98},
and the calculations of Giudice, Rattazzi, and Wells \cite{Giudice:1999ck}.  

The differential cross-sections for the parton processes relevant to 
graviton plus jet production in hadron collisions are given in 
Equations~\ref{sigf1},~\ref{sigf2}, and~\ref{sigf3}.

\begin{equation}
\frac{d^2\sigma}{dt ~dm}(q\bar q \to g G) = 
\frac{\alpha_s}{36}~
\frac{2\pi^{n /2}}{\Gamma (n /2 )}~
\frac{1}{s\md^{2+n}}~
m^{n -1}~ F_1 (t/s,m^2/s)
,
\label{sigf1}
\end{equation}

\begin{equation}
\frac{d^2\sigma}{dt ~dm}(qg \to q G) = 
\frac{\alpha_s}{96}~\frac{2\pi^{n /2}}{\Gamma (n /2 )}~
\frac{1}{s\md^{2+n}}~
m^{n -1}~F_2 (t/s,m^2/s)
,
\label{sigf2}
\end{equation}

\begin{equation}
\frac{d^2\sigma}{dt ~dm}(gg \to g G) = 
\frac{3\alpha_s}{16}~\frac{2\pi^{n /2}}{\Gamma (n /2 )}~
\frac{1}{s\md^{2+n}}~
m^{n -1}~F_3 (t/s,m^2/s)
\label{sigf3}
\end{equation}
The Mandelstam variable $t$ in Equations.~\ref{sigf1}-\ref{sigf3}
 is defined as $t=(p_q -p_G)^2$.

The calculation of graviton emission is based on an effective low-energy theory
that is valid at parton energies below the scale $\md$.

The $F_i(x,y)$ functions in Equations~\ref{sigf1}-\ref{sigf3} are
\begin{eqnarray}
F_1(x,y)&=&
\frac{1}{x(y-1-x)}\left[ -4x(1+x)(1+2x+2x^2)+ \right.
\label{eqf1}
\nonumber \\&& \left. y(1+6x+18x^2+16x^3)-6y^2x(1+2x)+y^3(1+4x)\right], \\
F_2(x,y)&=& -(y-1-x)~F_1\left( \frac{x}{y-1-x}, \frac{y}{y-1-x}\right) =
\nonumber \\ &&
\frac{1}{x(y-1-x)}\left[ -4x(1+x^2) +y(1+x)(1+8x+x^2)
\right. \nonumber \\ && \left.
-3y^2(1+4x+x^2)
+4y^3(1+x) -2y^4 \right], \\
\label{eqf2}
F_3(x,y)=&&\frac{1}{x(y-1-x)}
\left[ 1+2x+3x^2+2x^3+x^4
\right. \nonumber \\ && \left.
-2y(1+x^3)+3y^2(1+x^2)-2y^3(1+x)+y^4 \right] .
\label{eqf3}
\end{eqnarray}

The function $F_1(x,y)$ determines the cross-section for
$f \bar f \to \gamma G$.
$F_1(x,y)$ is
invariant under exchange of the Mandelstam variables $t$ and $u$;
this is reflected in the property $F_1(x,y)=F_1(y-1-x,y)$. The same
property holds also for the function $F_3(x,y)$, relevant to
the QCD process.

\begin{figure}
\centerline{
   \psfig{file=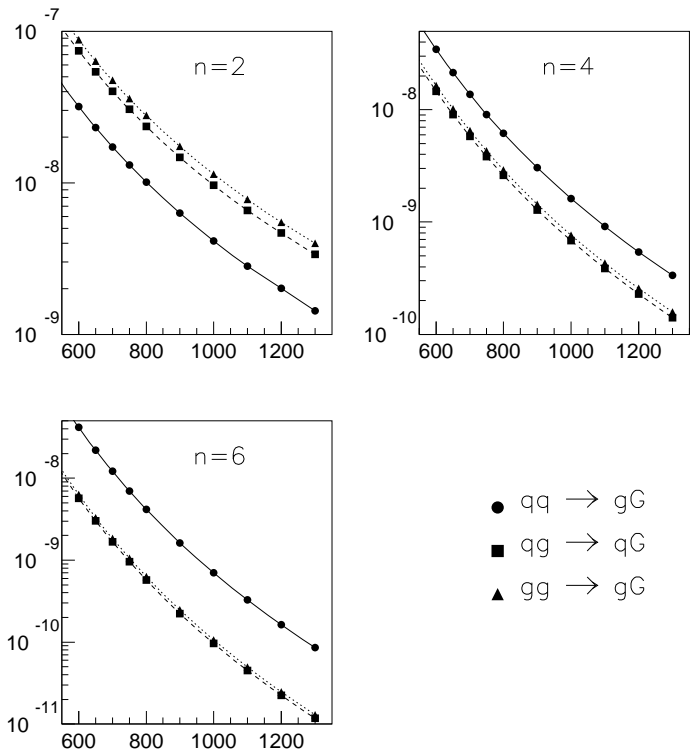,width=4.2in}}
  \caption{Cross sections for each subprocess for (a) n=2,
        (b) n=4, and (c) n=6.}
  \label{sigma_n}
\end{figure}

\begin{figure}
\centerline{
   \psfig{file=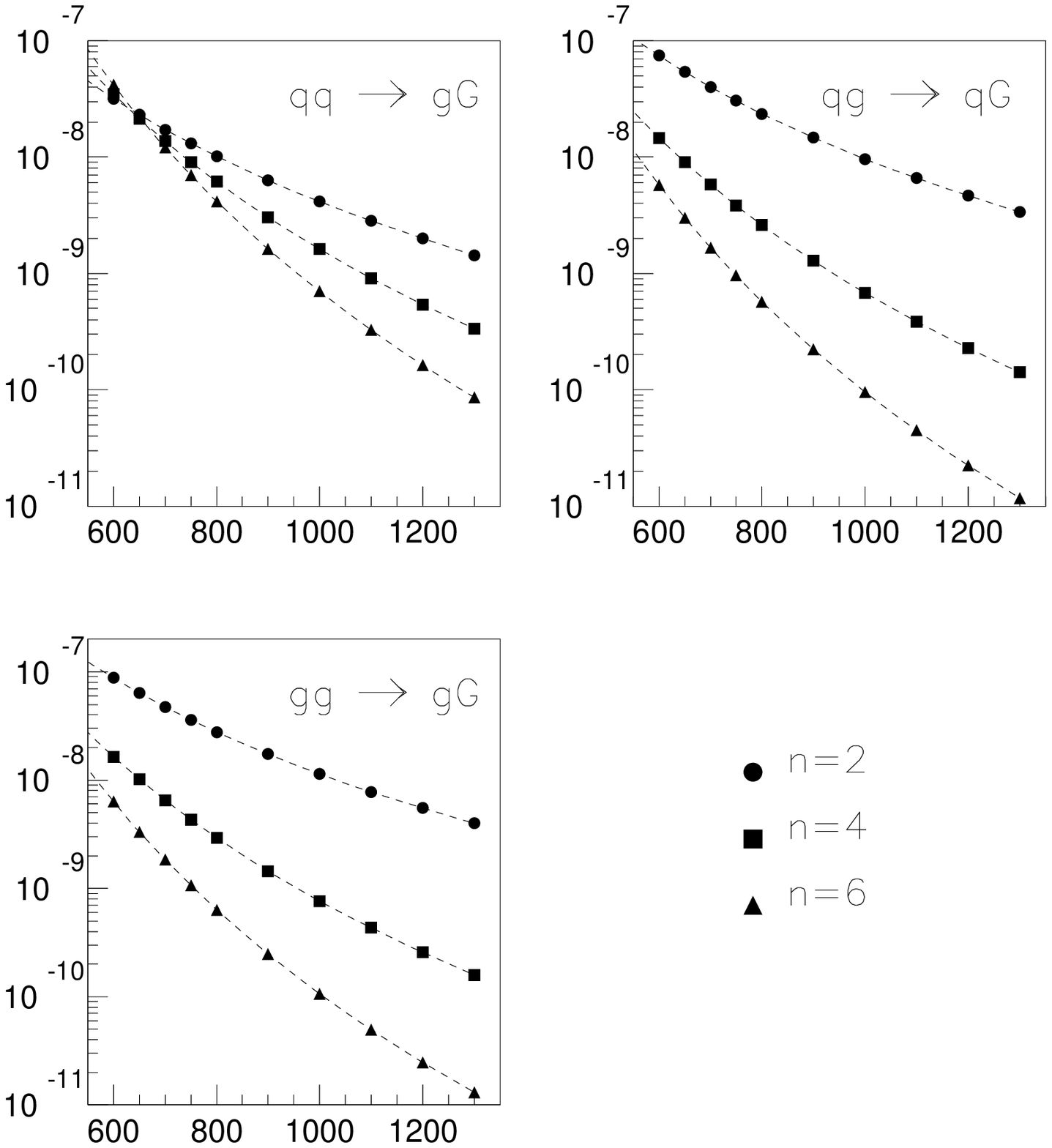,width=4.2in}}
  \caption{Cross sections for n=2, 4, and 6, for each subprocess:
        (a) $q\bar q \to g G$
        (b) $qg \to q G$, and 
        (c) $gg \to g G$.}
  \label{xvproc}
\end{figure}

Cross sections from \pyt\ for individual graviton production
sub-processes are shown
as a function of $M_D$ for different values of $n$ in Figures~\ref{sigma_n} 
and ~\ref{xvproc}.  Figure~\ref{sigma_n} compares the different 
sub-processes for the same value of $n$, while Figure~\ref{xvproc} compares
different values of $n$ for the same sub-process.

\begin{figure}
\centerline{
  \psfig{file=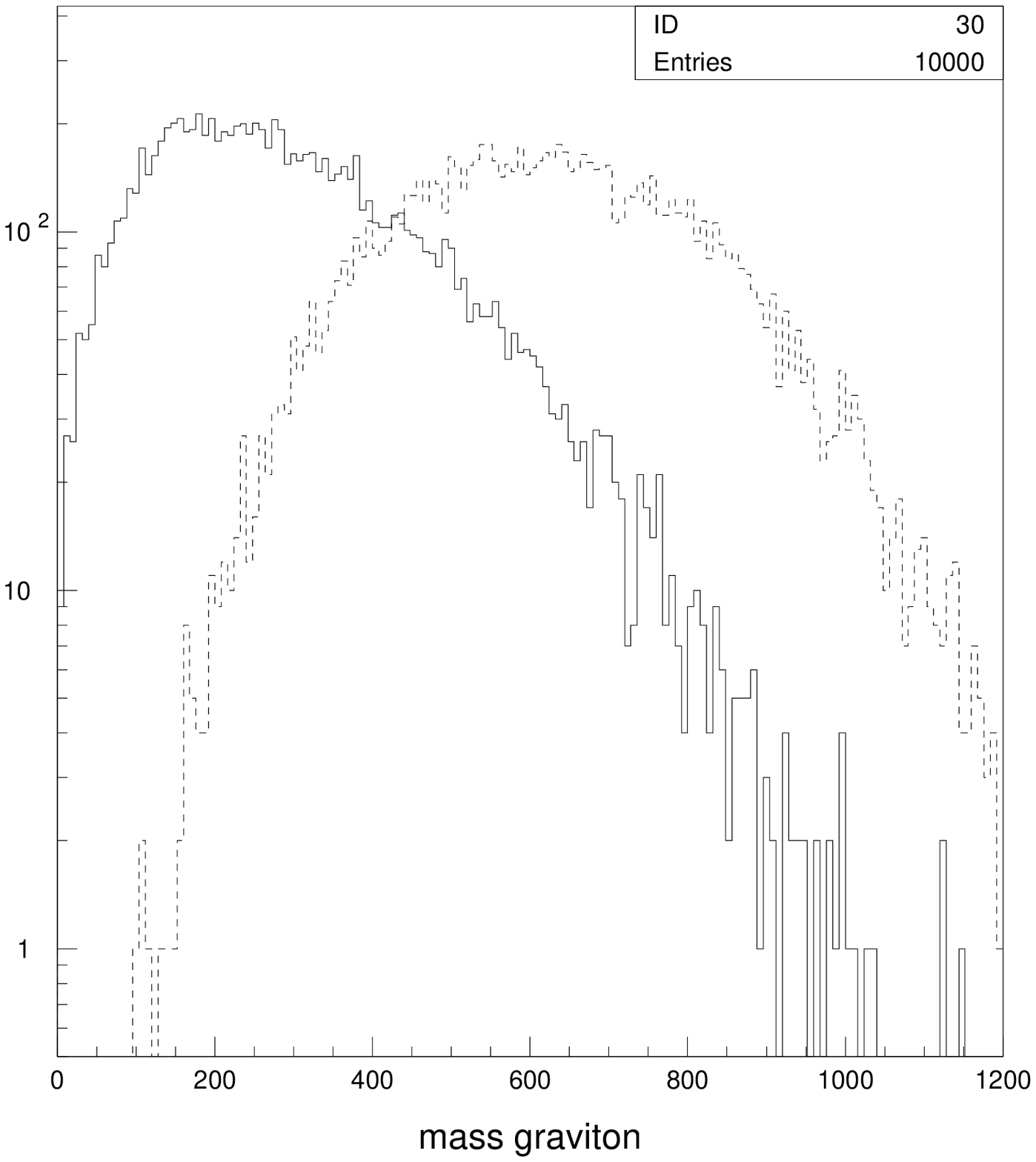,width=2.9in}
  \psfig{file=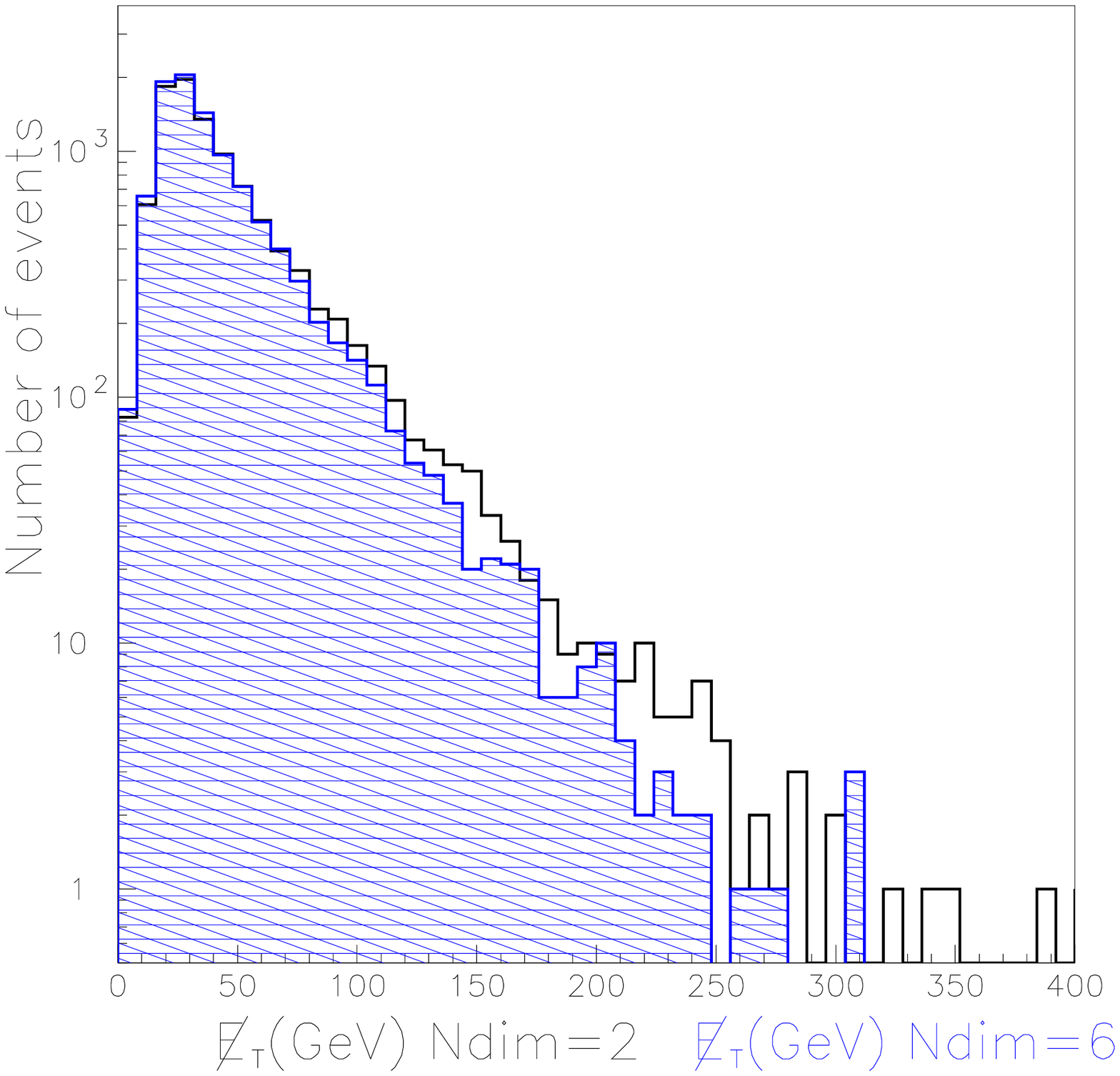,width=2.9in}}
  \caption{(l)The  graviton mass distribution 
           for $n=2$ and  $n=6$ extra dimensions.
(r) The \met\ distribution for n=2
           and  $n=6$  extra dimensions(shaded) for
           $qg \to q G$ process simulation.}
  \label{mass2_mass6}
\end{figure}

Note that the cross-section for $q\bar q \to g G$ is larger for larger
values of $n$, relative to the other sub-processes.  This can be traced
to the different dependences of $F_1,\ F_2,$ and $F_3$ on $m^2/s$
(labelled $y$ in equations~\ref{eqf1}-\ref{eqf3}).  $F_2$ and $F_3$ have a
dependence on quartic dependence on $y$, whereas $F_1$ has only a
cubic dependence.  This results in larger splittings at high values
of $M_D$ between different values of $n$ for  $qg \to q G$ and $gg \to g G$
compared to $q\bar q \to g G$.
For $M_D$ = 1 TeV a significant number of heavy KK gravitons will be
produced with masses averaging in the hundreds of GeV, as can be seen
in Fig.~\ref{mass2_mass6}.  The peak of the mass distribution is higher
for $n=6$ extra dimensions, because the density of KK states is a more
rapidly increasing function than for $n=2$.  This difference does
not show up in the \met\ distribution (Fig.~\ref{mass2_mass6}).  This is
due to two competing effects: (1) the heavier KK gravitons for $n=6$
have larger transverse energy, but (2) the rapidly decreasing parton
distribution functions cause the heavier gravitons to be produced near
threshold.  These two effects cancel, leaving nearly identical \met\ 
distributions for different values of $n$.

\subsubsection{The Analysis}
\label{chap:anal}
The data preselection
and the jet fiducial requirements are the same as in the case 
of the multijet plus \met\ analysis used in the search for gluinos and squarks and 
they are designed to retain a high purity 
in the real missing energy sample.  

In the missing energy plus one jet search,
the missing energy comes from the KK graviton tower
and the one jet from the recoiling parton.
To reduce the  background contribution from $W\ra\ell\nu$+jets 
we apply the same   {\it indirect lepton veto} as in the multijet analysis.

To define the signal region we use three variables:
the \met~, the $N_{jet}$ and the Isolated Track Multiplicity \niso.
The requirements are, no isolated tracks, large missing energy and one or two 
jets (when two the second must not be grossly mismeasured).

The value of the missing energy requirement for the definition 
of the graviton signal region
is driven by the missing energy Level 2 trigger efficiency 
 and optimized for the graviton signal. 
The $N_{jet}$ requirement
is motivated by the graviton monojet signal characteristic final state. 
The second jet is primarily allowed so that the QCD background can 
be calculated using the HERWIG Monte Carlo and the CDF detector simulation 
with a reliable normalization from the CDF QCD data,  and additionally  
so that the systematic uncertainty of the signal due to ISR/FSR
is kept as low as possible. 
Allowing a second jet also permits interpretation 
of the results with a k-factor inclusion in the signal cross section 
estimate.

The \niso~  requirement increases 
the sensitivity to the graviton 
signal by reducing the $W$+jet contribution,
while at the same time retaining the signal.
The rest of the analysis path is based on the kinematics 
and aims at high sensitivity for the signal.

\subsubsection{Summary of \sm~ processes with \\$\met~+$ jet  in the  final state}
The \sm\ processes with large missing energy and multijets in the 
final state that constitute backgrounds to the graviton search are

\begin{itemize} 
\item $Z$+jets :  QCD associated $Z$ production with   
$Z \rightarrow \nu \nu$  is the most significant and irreducible 
background component. 
For the $Z+$jets backgrounds the \pyt~ Monte Carlo simulation 
is used and normalized to the $\Zee +\ge 1$ jet data (our standardizable candle).
 
\item $W+$jets :  QCD associated $W$ production with  $\Wen$, 
$\Wmn$, and $\Wtn$ has a large \met\  + jet contribution.    
For the $W+$jets and backgrounds the \pyt~ Monte Carlo simulation 
is used and normalized using the $\Zee +\ge 1$ jet data sample.

\item top, single top : In $t\bar t$ production a $W$ from a  top decay 
can decay semileptonicaly and contribute to the high \met~ tails.  We use 
\pyt~ to simulate \ttb~ production with all inclusive top decays.   For the 
normalization to the data luminosity the theoretical calculation 
$\sigma(t\bar t)=5.1 ~{\rm}pb~\pm~ 18\%$ (which is consistent with the CDF  
$t \bar t$ \cite{tdib} cross section) is used.  
\herw~ and \pyt~ Monte Carlo programs are used to simulate single 
top production via W-gluon fusion and W*  production respectively. 
The theoretical cross sections   $\sigma(tq^{\prime})=1.7~ {\rm pb}~\pm 15\%$ 
and   $\sigma(bt)=0.73 ~{\rm pb}~\pm 9\%$ ~\cite{tdib} are used to normalize 
the samples.  

\item dibosons : For $WW$, $WZ$, $ZZ$ production  \pyt\ is used.
For the normalization, the theoretical cross sections calculated for each 
diboson process are used: 
$\sigma(WW)=9.5\pm 0.7 ~{\rm pb},~ \sigma(WZ)=2.6\pm 0.3 ~{\rm pb}$ and 
$\sigma(ZZ)=1 \pm 0.2 ~{\rm pb}$ ~\cite{tdib} are used.

\item  QCD : The  QCD background is generated with  \herw~ and normalized
 to the CDF jet data using dijet events. 
Clearly one does not expect this to be 
a significant physics background.
\end{itemize}

Once the signal to background ratio is optimized as a function of the measured variables
(kinematical,topological etc), the requirements are set and the data are compared 
to the standard model predictions. From there of course we can interprete 
the results as a limit in the allowable cross section of non-standard model processes,
extra dimensions and what not. The result of this particular analysis has been reported in conferences and will appear
soon in the literature but similar analyses from CDF and D0 have already reported results \cite{other}.

\section{What next?}

It is a very interesting time for physics, indeed for fundamental physics.
Recent experimental cosmology results are changing the ideas we
had about how much of the universe we know and how well. Results
in particle physics, such as the massiveness of the neutrinos, 
the tremendous precision at which the Standard Model holds in the 
up to now explored electroweak scale region (without the Higgs boson
showing up), are all puzzling us. Putting together a picture of 
how spacetime happens and how it is filled, 
by means of observations and studies
in a controlled experimental environment, is the primary and urgent 
work of both experimentalists and theorists. In these lectures I gave
examples of how we go about this using colliders. 

Many thanks to the school organizers Joe Lykken, Steve Gubser, and Kalyana T. Mahanthappa; the TASI 2001 students; Kevin Burkett and all my CDF collaborators always.  
The author is supported by NSF and the Pritzker Foundation.

\addcontentsline{toc}{chapter}{Bibliography}

\begin{thebibliography}{20}
\bibitem{multiple_vertices} If there are multiple vertices in the
event we use the vertex with the largest $\sum P_{T}$ of associated tracks.
\end{thebibliography}
\end{document}